\documentclass[reqno]{amsart}
\usepackage{graphicx}
\usepackage{epsf}
\usepackage{color}
\newcommand{\nc}{\newcommand}
\nc{\tcb}{\textcolor{blue}}
\nc{\tcr}{\textcolor{red}}
\nc{\tcg}{\textcolor{green}}
\def\sigmabf{\mbox{\boldmath $\sigma$}}
\def\bvec{\mathbf}
\def\be{\begin{equation}}
\def\ee{\end{equation}}
\def\ba{\begin{array}}
\def\ea{\end{array}}
\def\bea{\begin{eqnarray}}
\def\eea{\end{eqnarray}}
\def\drm{{\mathrm d}}
\def\erm{{\mathrm e}}
\def\dis{\displaystyle}
\newcommand{\grbf}{\mbox{\boldmath $\nabla$}}
\def\la{\grbf^2 \,}

\def\ov{\overline}

\begin{document}

\title{The genesis of the quantum theory of the chemical bond}

\author{S. Esposito}
\address{{\it S. Esposito}: I.N.F.N. Sezione di
Napoli, Complesso Universitario di M. S. Angelo, Via Cinthia,
80126 Naples, Italy ({\rm Salvatore.Esposito@na.infn.it})}%

\author{A. Naddeo}
\address{{\it A. Naddeo}: Dipartimento di Fisica ``E.R. Caianiello'',
Universit\`a di Salerno, and CNISM Unit\`a di Ricerca di Salerno, Via Ponte don Melillo, 84084 Fisciano (Salerno), Italy
 ({\rm naddeo@sa.infn.it})}%

\begin{abstract}
An historical overview is given of the relevant steps that allowed the genesis of the quantum theory of the chemical bond, starting from the appearance of the new quantum mechanics and following later developments till approximately 1931. General ideas and some important details are discussed concerning molecular spectroscopy, as well as quantum computations for simple molecular systems performed within perturbative and variational approaches, for which the Born-Oppenheimer method provided a quantitative theory accounting for rotational, vibrational and electronic states. The novel concepts introduced by the Heitler-London theory, complemented by those underlying the method of the molecular orbitals, are critically analyzed along with some of their relevant applications. Further improvements in the understanding of the nature of the chemical bond are also considered, including the ideas of one-electron and three-electron bonds introduced by Pauling, as well as the generalizations of the Heitler-London theory firstly performed by Majorana, which allowed the presence of ionic structures into homopolar compounds and provided the theoretical proof of the stability of the helium molecular ion. The study of intermolecular interactions, as developed by London, is finally examined.
\end{abstract}

\maketitle

%keywords: Ettore Majorana, chemical bond, homopolar compounds, ionic structures

%--------------------------------------------------------------------------
%--------------------------------------------------------------------------

\section{Introduction}
\label{intro}

\noindent The history of atomic physics is inextricably related to that of the quantum theory, as well known, since the development of the last one was required just by evidences claiming for an appropriate physical description of atomic systems. If it is true that the first success of the old quantum theory relied mainly in N. Bohr's characterization of the hydrogen atom, the general achievement of the new quantum mechanics was driven by the accurate description of the helium atom \cite{nostro}. Predictions about heavier atoms involved, instead, just the development of appropriate (mathematical and physical) methods to be applied within a well established conceptual framework: in a sense, it was only a matter of numerical accuracy. A quick glance to the monumental work by Mehra and Rechenberg \cite{MR} may serve very well to seize the path followed by atomic physics as quantum mechanics took its primary (and secondary) steps, and the conclusion that can be drawn is simply that, irrespective of more or less expected difficulties, such a path was somewhat linear to a very large extent.

The same conclusion cannot be applied at all to the development of molecular physics, whose path within quantum mechanics was far more intricate, at least at its starting, ranging approximately from 1926 to 1931. The reason for this resides mainly in the basic difference between atomic and molecular systems, i.e. to deal with a two- or more-center problem, with respect to the one-center problem for atoms, but it is only partially true that the sole greater mathematical complexity was at the root of the intricacy of that path. If polar molecules were described simply in terms of electrostatic interaction among the constituent atomic systems already in the framework of the old quantum theory, novel physical concepts were required even just for an approximate understanding of the chemical bond in homopolar compounds. And, although such concepts revealed to be genuinely quantum mechanical in nature, their emergence and physical interpretation was not at all trivial, as testified by the appearance (in the years mentioned) of a number of papers discussing different methods and viewpoints.

As a matter of fact, there was not a unique line of research to which subsequent results were added as they were got, in order to obtain a more and more clear picture, but, rather, several lines appeared that produced different (and, in some cases, competing) results without leading to a univocal advancement in the clear understanding of the issue of the chemical bond. Conversely, also several minor contributions revealed their relevance in making the general picture clearer, and the result of this marked intricacy was that, only in the first years of 1930s, the multifarious facets of quantum chemistry came out limpidly.

The concept of a {\it chemical bond} made its first appearance in the chemical literature in 1866, in a paper by E. Frankland \cite{Frank}, and, as well, the idea of {\it valence} was very early introduced into chemistry to explain some number relationships in the combining ratios of atoms and ions. However, the first attempt to incorporate the atomic structure information in a consistent -- though qualitative -- theoretical framework was performed by G.N. Lewis only in 1916 \cite{Lewis}, the key points of his theory being that each nucleus tends to be surrounded by a closed shell of electrons (as in noble gases) and that the homopolar chemical bond is built from a pair of electrons shared between two nuclei.

Despite some success in describing a number of previously unrelated facts, Lewis theory was not able to say anything on the nature of the forces involved in the formation of the homopolar bond \cite{history1} and, on the other hand, Bohr's old quantum theory, with its successful explanation of Balmer's law, triggered the key role of spectroscopy as a guiding principle in explaining also the structure of the chemical bond. Nevertheless, only with the advent of quantum mechanics appropriate and powerful theoretical tools became available in order to tackle the problem \cite{Mehra}, thus becoming possible -- in principle -- to write down an equation for any system of nuclei and electrons, whose solution would have provided thorough predictions on the stability of the molecular system. However, the $n$-body problem revealed to be not amenable to exact analytical solutions, thus triggering the development of several approximation methods, and the key idea of {\it exchange forces} (or quantum {\it resonance}), introduced in molecular physics by W. Heitler and F. London (and borrowed by W. Heisenberg's theory of  the helium atom \cite{Hei26}), was finally required in order to lay the foundations of the quantum theory of homopolar chemical bond.

As happened already in atomic physics, spectroscopy also contributed as a key tool towards the development of such a theory, and several scientists, including F. Hund, R.S. Mulliken, G. Herzberg and  J.E. Lennard-Jones, appealed to molecular spectroscopy in order to introduce the concept of {\it molecular orbital}, which produced an approach to the understanding of the chemical bond completely different with respect to that introduced by Heitler and London. A number of different refinements and generalizations of both approaches later appeared in the subsequent literature, but quantitative calculations remained much too complicated to allow tests of the novel ideas in molecules other than diatomic ones formed from hydrogen and helium atoms.

Different approximation methods were, then, developed (even for the simplest molecules), and some physical insight into the complex mathematical problem was gained thanks to the seminal paper by M. Born and J.R. Oppenheimer. Here it was shown that, at a first approximation, the motion of the nuclei in the molecules could be neglected, and a well-defined hierarchy existed between electronic, vibrational and rotational states.
Additional light upon the intricate problem came, then, also from the application of the powerful tools of group theory, although only few authors exploited it appropriately to get some insight into the mathematical form of the wavefunction, deduced from the symmetry properties of the molecular system it describes. To a larger extent, group theory was used instead to put some order in spectroscopic evidences, and to relate apparently different observations.

The transmission of the achievements gained by the quantum theory of the chemical bond to people more accustomed to a chemistry-based language produced, remarkably, also a non-negligible result in the understanding of several points of the theory itself. L. Pauling and J.H. van Vleck, indeed, in addition to their own original contributions to the subject, wrote some review papers that served to clarify substantially the physical meaning of the novel concepts introduced, and thus helped to digest the corresponding mathematical results, both to chemists and physicists.

In the present paper, we will try to outline the relevant steps that allowed the genesis of the quantum theory of the chemical bond through the years 1926-1931, being well aware that, as a consequence of what explained above, such a work is {\it de facto} partial and biassed, if it is contained -- as we do -- into a reasonable length. Nevertheless, the discussion of general ideas as well as some important details in molecular spectroscopy (Sect.\ref{spectrosect}) and quantum computations, both perturbative and variational, for simple systems (Sect.\ref{bosect}), along with a thorough description of the historical development of the method of molecular orbitals (Sect. \ref{molorb}) and that of the Heitler-London theory (Sect.\ref{hlsect}), can certainly serve as a starting point in the comprehension of that genesis. A completely non-negligible role in such a task was played also by the results about the simplest molecules formed with hydrogen and helium, discussed in Sect.\ref{further}, as well as those pertaining to the subject of intermolecular interactions, treated briefly in Sect.\ref{intermol}. We will focus mainly on diatomic molecules, as indeed happened in the years considered for quantitative studies. The large amount of contributions here analyzed will be finally summarized in our concluding section.

\section{Spectroscopy of diatomic molecules}
\label{spectrosect}

\noindent Early days investigations about simple diatomic molecules were aimed mainly at giving a qualitative explanation of the structure of the spectra observed, just by applying phenomenological models already tested in the interpretation of atomic spectra. The simple idea followed at the very beginning \cite{spectra1} was to ascribe all the known phenomenology to either the rotation or the vibration motion of the nuclear component of the molecules, but it was realized quite soon \cite{spectra2} that a fine structure originated from electronic motion, thus strongly influencing the observed band spectra. The theoretical problem then arose, within the framework of the new quantum mechanics, to explain the relationship between the electronic motion and the band spectrum.

\subsection{Molecular spectroscopy derived from atomic}

A key contribution in this direction came from a seminal work by Hund \cite{hund2a}, who introduced the concept of an {\it adiabatic} combination of two atoms in forming a molecule through the application of a vector addition model, resulting into a first classification of the spectral terms of diatomic molecules.

His starting point was the fact that the state of an atom with one valence electron is determined by  the four quantum numbers $n$, $\ell$, $j$, $m$, where $\ell$ labels the orbital angular moment of the electron, $j=\ell+s$ is the total angular momentum of the atom ($s$ being the electron spin), and $m$ is its component along the direction of an external field. The counting of the number of terms in an atom with several outer electrons (labelled by the index $r$), instead, followed from the general assumption that the motion is essentially determined by the principal and secondary quantum numbers of each electron: the interaction among the vectors $\ell_{r}$ and among the vectors $s_{r}$ is taken into account as a first-order perturbation, while the interaction of $\ell$ and $s$ is a second-order effect. As a result, the correct counting of terms came out, but the position of terms in the spectrum was wrongly predicted, since the correct ratio between the different interactions was no longer kept.

Hund applied such results just to the study of diatomic molecules, by assuming -- in the first instance -- the nuclei to be fixed in their positions by some external force. In the presence of only one electron, the state of the molecule could then be determined by the quantum numbers $n$, $\ell$, $j$, $m$, with different $m$-values resulting in different terms, even without an external field. In particular, if $i$ denotes the angular momentum along the direction of the internuclear axis, and the perturbation induced by the two nuclei on the electronic motion is large with respect to the interaction between $\ell$ and $s$, the actual positions of the terms resulted to be well described by the numbers $n$, $\ell$, $i_{\ell}$ and $i_{s}$, with $ i_{\ell}\leq \ell$ and $i_{s}=\pm {1}/{2}$. Instead, when several electrons are present, Hund considered a model of a molecule as built up from joining together two atoms (described by quantum numbers $\ell_{1}$, $s_{1}$, $j_{1}$, $i_{1}$ and $\ell_{2}$, $s_{2}$, $j_{2}$, $i_{2}$) with the required total number of electrons, in such a way that the corresponding molecular term could be determined by the configuration labelled by $\ell_{1}$, $\ell_{2}$, $s_{1}$, $s_{2}$, $j_{1}$, $j_{2}$, $i_{1}$, $i_{2}$. An alternative description to the ``united atom'' was, as well, introduced by imagining the molecule to be formed by only one atom with the required number of electrons, whose nucleus is then split and the resulting parts taken a little apart: in such a way, the molecular terms would be determined by $\ell$, $s$, $j$, $i$. By following W. Pauli \cite{pauli1}, Hund finally concluded that the two alternative countings led to the same number of terms in the molecular spectrum.

The next step was to relax the assumption of fixed nuclei, and let them to perform a motion such that the corresponding energy is small when compared to the electronic ones (in the limit of large nuclear masses). By
keeping the internuclear distance to be fixed, the motion of the molecule would have been determined by two additional (with respect to the electronic) quantum numbers, $p$ and $q$, identified with the total angular momentum and its component along the direction of the external field, respectively, as well as the vibrational quantum number.

In order to estimate the position of the molecular terms, Hund assumed, at a first approximation, that the electrons circled around both nuclei, and the following interactions were included: 1) coupling among the angular momenta $\ell_{r}$ and among the spins $s_{r}$ of single electrons; 2) perturbation induced by the motion of the two nuclei on the electronic motion (``\textit{influence of the distinguished axis}''), determined by $\ell_{r}$ and independent of the direction of the motion around the internuclear axis; 3) coupling between $\ell_{r}$ and $s_{r}$; 4) influence of $\ell_{r}$ on the molecular rotation through a change of its moment of inertia. The evaluation of the \textit{influence of the distinguished direction of the internuclear axis}, which resulted to be very similar to the Stark effect within Hund's approximations, was performed by considering constant the component of the total angular momentum of the electronic orbit around the nuclear axis, and thus quantized to first order with the (small) perturbations induced by interactions 3) and 4).

Two limiting cases were, then, considered, corresponding to two particular mo\-le\-cu\-le configurations -- Hund's  a) and b) configurations -- where rotational effects are small or large with respect to those driven by the $\ell_{r} \! - \! s_{r}$ coupling.

In case a), a rapid electronic motion was obtained as a first approximation, with a corresponding angular momentum along the nuclear axis equal to $i_{l}$, while the total vector $s$ was found to adjust to the vectors $\ell_{r}$ (in second approximation) and to perform a precession around the internuclear axis. This gives rise to $2s+1$ terms with $i=i_{l}+s,i_{l}+s-1,...,i_{l}-s$. The effect of the rotation was finally added, producing the quantization of the total angular momentum $p$,
%as expressed in the rotational energy expression $\frac{h^{2}}{8\pi ^{2}A} p\left( p+1\right)$-const,
along with a splitting of the single-electron terms in the presence of a slow rotation.

In case b), instead, the rotation is added to the electronic motion as a second order effect,
%and an energy $\frac{h^{2}}{8\pi ^{2}A} p_{\ell}\left( p_{\ell}+1\right)$-const is obtained at a first step.
while, as a third order effect, a perturbation of this motion induced by $s$ is included, so that both $p_{\ell}$ and $p_{s}$ precess around a total momentum $p$. As a result, the energy is proportional to $\overline{\cos (\ell \cdot s)}$ and, in the case of a fast rotation, a splitting of the electronic terms in $2s+1$ components is obtained.

By comparing cases a) and b), Hund finally found that the same values of $p$ were obtained, the corresponding band lines being explicitly depicted.

\subsection{Adiabatic transitions}

In a series of other papers \cite{hund2b} \cite{hund2c}, Hund further developed his qualitative model of a diatomic molecule and also generalized it to a molecule with an arbitrary number of atoms, by focusing on the characteristics of the spectral terms related to the electronic motion (so that molecular optical spectra were considered, where the motion of the outer electrons is crucial).

A classical model of a diatomic molecule dealt with the motion of a charged mass point, identified with the optical electron, in a central force field with a potential with two minima, $U=U_{1}\left( r_{1}\right) +U_{2}\left( r_{2}\right) $, $r_{1}$ and $r_{2}$ being the distances of the given electron to the nuclei. This problem was worth to be solved in two limiting cases. The first case corresponds to a very large distance between the nuclei, so that the electron is in the neighborhood of the first or the second nucleus, respectively: a {\it resonance} \cite{Hei26} \cite{hei1} between the two configurations can be envisaged, and the motion is similar to that of an electron in an atom placed in an external electric field. The second one corresponds, instead, to a situation in which the nuclei are close together, and the motion of the electron can be described by a simple perturbation theory in a modified one-minimum potential. The actual configuration of a diatomic molecule pertains, obviously, to an intermediate case, but, as pointed out by Hund, the transition from one limiting case to the other cannot be done adiabatically within such a classical model.

Hund then switched to the quantum version and obtained an {\it adiabatic relation} between the states of the two separated atoms or ions, those of the corresponding diatomic molecule and the states of the atom resulting from the union of the two nuclei. Such relation allowed him to solve the general two-center problem and to build up a reliable model system giving rise to a qualitative explanation of molecular spectra.

Indeed, for the simplest two-center problem with one electron, he obtained a Schr\"odinger equation which is
separable in elliptic coordinates $\xi $, $\eta $, $\varphi $, as earlier derived by W. Alexandrow \cite{alexandrow} and \O. Burrau \cite{burrau} (see below), but getting a more general solution describing the qualitative behavior of the corresponding molecular terms. For a molecule with more than two electrons, Hund considered more than two separated systems; for diatomic molecules with two electrons, the systems are those with both electrons on one nucleus, both electrons on the other nucleus, and one electron on each nucleus. He worked out again two limiting cases, the most relevant one assuming a very weak coupling between the electrons, so that the corresponding Schr\"odinger equation is approximately separable and then amenable to find solutions for the motion of each single electron.

Hund applied such considerations to the study of the simplest molecule, i.e. the hydrogen molecular ion H$_{2}^{+}$, and built up a model taking into account the vibration of the internuclear distance around an
equilibrium configuration, determined previously by the electronic configuration. In this picture, the energy becomes a function of the internuclear distance $r$ and its minimum gives the equilibrium configuration. In the limit of small $r$, the energy of each term goes as ${1}/{r}$ and the distances between terms behave as those between terms of a \textit{united nucleus}.

A further application of the theory was to the hydrogen molecule H$_{2}$, whose energy term scheme was recognized to lie on the boundary between the systems H$+$H, H$^{+}+$H$^{-}$ and the He atom, in this
way opening the door to the concept of a {\it polar molecule}:
\begin{quote}
a molecule is polar if it transforms into two oppositely charged ions when the nuclei are pulled apart \cite{hund2b}.
\end{quote}

Finally, following Oppenheimer \cite{oppi1}, a generalization to molecules with four or more atoms was as well developed \cite{hund2c}, starting from a picture where the effect of the electronic motion is large with respect to nuclear vibrations, which in turn are large if compared with the rotation of the molecule. Thus, only vibrational and rotational spectra were considered, and the symmetry characters of particular configurations were derived, such as -- for example -- those of molecules with two different ``arrangements'' of minimal potential energy which are equivalent up to a reflection: the transition from one arrangement to the other corresponds to a given frequency in the spectrum.

As summarized later by E.U. Condon,
\begin{quote}
the analysis of Hund provides the important result that the electronic term of the lowest state of a molecule changes continuously from its value for a neutral atom of equal number of electrons to its value for the dissociated atoms, according to the new quantum mechanics \cite{condon2}.
\end{quote}

%The physical picture for the process of combination or separation of atoms at the band convergence point was further developed experimentally by Franck \cite{franck1} and theoretically by Condon \cite{condon1}, soon followed by Kronig \cite{kronig1} who clarified some issues in the work by Hund.

\subsection{Quantum-mechanical computations}

In 1927, a general formulation of the complete quantum-mechanical problem of a diatomic molecule in terms of a partial differential equation with $3N+6$ independent variables, $N$ being the number of electrons, was barely considered by Condon \cite{condon2}. A two-center problem was usually solved by replacing the given system with a simpler one, built of two masses interacting each other through an arbitrary potential energy law. This effective potential energy was identified with the average of the interactions between the rapidly moving electrons and the nuclei, in addition to the mutual Coulomb repulsion between nuclei. Condon then argued that, due to the heavy masses of nuclei, an approximate separation of variables in the Schr\"odinger equation could be performed, in this way anticipating some concepts which should have been developed some months later by Born and Oppenheimer \cite{bo1} (see below).

He applied such considerations to the hydrogen molecular ion, analyzing two limiting cases. In the first case, when the nuclei are far apart, the system reduces to a free hydrogen atom plus a proton, so that the electronic energy is mainly given by the Coulomb interaction between the proton and the electronic cloud around the atom. Conversely, when the nuclear separation is zero, the electron moves in the field of a double central charge and its energy coincides with that of the lowest state of ionized helium.

The approximate solution of the neutral H$_{2}$ molecule, then, easily followed by assuming each of the two electrons to move independent of the other in the ground state of the H$_{2}^{+}$ system, as given by Burrau calculations \cite{burrau}. Within this simple model Condon \cite{condon2} obtained an electronic energy twice as that of H$_{2}^{+}$ and a definite lower limit on the moment of inertia of the molecule. Furthermore the electronic interaction resulted to be always positive and to decrease with the nuclear separation. Earlier empirical results by R.T. Birge and H. Sponer \cite{bsponer} about the dissociation of unexcited molecules into two unexcited atoms, as well as related theoretical considerations by Hund \cite{hund2b}, were thus recovered and generalized.

In close agreement with the experimental findings by J. Franck \cite{franck1}, Condon's analysis also led to the result that the motion of the nuclei in a molecule gets modified by electronic changes through their influence on the molecular binding. This paved the way to further developments of the quantum theory of the chemical bond as due to electrons, which ended up a few months later with the seminal contributions by Mulliken \cite{mulliken2} \cite{mulliken3}, Heitler and London \cite{heitler1}, and Pauling \cite{pauling1}.

\subsection{Peculiarities}

A theoretical explanation of some intriguing features in the band spectra of diatomic molecules was given by R.L. Kronig \cite{kronig1} by means of approximation methods (perturbation theory) for the solution of the Schr\"odinger equation, together with some considerations on the symmetry properties of the eigenfunctions. He realized that the electron angular momentum is quantized in its direction parallel or antiparallel to the internuclear axis, predicting the corresponding doublet splitting labelled by the rotational quantum number. He also obtained an interesting interpretation of irregularities in the term structure (called {\it perturbations}), as due to particular combinations of electronic, vibrational and rotational quantum numbers. Finally a description of the phenomenon of predissociation, discovered by Henri \cite{henri} in 1924 and later investigated by G. Wentzel \cite{wentzel} and E. Fues \cite{fues1}, was given together with an estimate of the lifetime of the predissociated molecule. The dissociation of a molecule into two separated atoms was found to take place by increasing the vibrational energy for a given electronic state; starting from a series of discrete vibrational states, continuous eigenvalues could be reached:
\begin{quote}
energetically the possibility exists for the molecule to dissociate radiationless into atoms, by a transition from the first to the second electronic state \cite{kronig1}.
\end{quote}
In a companion paper \cite{kronig2}, the spontaneous decomposition of a diatomic molecule obtained by increasing the rotational energy was discussed as well, by resorting to a simple model first introduced by R.W. Gurney and Condon for the interpretation of radioactive decay \cite{gurney}. A theoretical explanation of previous experimental findings on the normal state of the HgH molecule \cite{hult2} was also supplied.

The first clear evidence of the existence of electronic levels in addition to vibrational and rotational ones had to be recognized in the He$_{2}$ spectrum, as pointed out by W.E. Curtis \cite{curtis1} and W. Weizel \cite{weizel1}. On the basis of Hund's findings \cite{hund2a} \cite{hund2b} \cite{hund2c}, they realized that the non-central field of the two separated nuclei could give rise to further term sequences in addition to those already known in atoms, a novel quantum number appearing as a result of the coupling of the electronic orbital angular momentum (quantum number $\ell$) with that associated to the internuclear axis, whose values run from $0$ to $\ell$ by integer steps. Weizel \cite{weizel1} also provided an explanation of some anomalies of the rotation term differences by postulating a change of coupling of the electronic orbital angular momentum driven by an increase of the rotational angular momentum. The anomalous appearance, as well as the intensities of the branches of some bands in the He$_{2}$ spectrum, were reported also by G.H. Dieke \cite{dieke}, who pointed out how all the intermediate stages between Hund's cases a) and b) could be observed in such a spectrum. His results were recovered and generalized by Curtis and A. Harvey \cite{curtis2}: in each electronic sequence, rotational terms were found to tend always to the same set of limiting values, which had to be identified with those characteristic of the He$_{2}^{+}$ ion.

\subsection{Group-theoretical methods}

Further contributions in molecular spectro\-sco\-py, as a tool for understanding the chemical bond, were provided by E. Wigner and E.E. Witmer \cite{WW1}, who introduced group-theoretical methods in studying several issues on the structure of molecular spectra. They again considered diatomic molecules (both with identical and with different nuclei), building up appropriate first order eigenfunctions just on the basis of symmetry properties.  Selection and intensity rules between the different types of terms, as well as the {\it aufbau} rule of a rotational band, were derived by assuming, in a first instance, fixed nuclei and then considering Hund's two cases, where the rotational energy is larger (case b) or smaller (case a) than the energy of the multiplet splitting. Molecules formed with light elements, such as H$_{2}$ and He$_{2}$, were found to be examples of the first case, while heavier molecules, such as I$_{2}$ and Hg$_{2}$, were recognized to fall in the second case, the intermediate region being instead quite small, due to the decrease of the rotational splitting  and the increase of the multiplet splitting with the atomic number. The aufbau rule of the electronic terms of a molecule followed from their detailed investigation on the relations between atomic and molecular spectral terms coming out when two atoms are ``united" into a molecule (following Hund's reasoning).

The basic procedure adopted by Wigner and Witmer was to evaluate, as a first step, the \textit{group-theoretically possible terms} of a given molecule, resulting from the two given families of atomic terms when the atoms are brought together to form the molecule. In such a way, they considered all terms for which the energy corresponding to large vibrational quantum numbers converges to the energy values of the separated
atoms, and, of course, such terms are much more in number than those actually realized. Also, they included
``limit" molecular terms that reduce to atomic terms when the two nuclei are united, provided that all the symmetry properties are preserved. The way they devised to remove, among the ``possible" terms, those belonging to the continuum spectrum (which are not observed experimentally), was to consider deep-lying atomic terms corresponding to the atom obtained from the union of the two nuclei.

Previous results by London \cite{london2} were, thus, recovered: only one molecular term was found to belong to the discrete spectrum, giving rise to atoms in their ground states at high vibrational quantum numbers, while the other one pertained to the continuum. The case of the molecule formed with an excited atom plus a normal one was allowed provided that two states with total quantum number equal to $2$ of the first atom combine with the ground state of the other, all other combinations being forbidden by selection rules, in agreement with Heitler and London findings \cite{heitler1} (see below). The aufbau rule was, then, successfully applied to the spectra of He$_{2}$ and Na$_{2}$ (and, to a lesser extent, to O$_2$), and really produced a classification of all the possible molecular terms, with a sensible advantage with respect to Hund's work:
\begin{quote}
The method used by us has instead the advantage that the pulling apart of the nuclei represents a process of which the course can be read from the convergence of the band spectra, so that the correspondence between the terms of the molecule and the terms of the separated atoms is given uniquely by the spectra \cite{WW1}.
\end{quote}
The only drawback of this approach was that nothing could be said about the stability of terms obtained: such a problem was discussed, instead, by Mulliken \cite{mulliken1} \cite{mulliken2} \cite{mulliken3}, as we
will see in Sect.\ref{molorb}.

Group-theoretic methods were also employed by other authors, in order to give a thorough theoretical interpretation of further experimental findings in molecular spectra.

E. Hill and van Vleck \cite{hill1} focused on the effect of the molecular rotation on spin multiplets, by computing amplitude matrices corresponding to Hund's cases a) and b), and remarkably providing a detailed treatment of the intermediate case. Indeed, this was carried out starting from case b) and introducing a coupling energy proportional to the cosine of the angle between the axis of electronic (total) spin $s$ and the molecular axis: $H_{\rm pert}=A \, \sigmabf_{k}\cdot {\bf s}$, where $\sigmabf_{k}$ is the component of electronic angular momentum along the symmetry axis of the molecule. Such a perturbation term was typical of the interaction of a spinning electron moving in a Coulomb field, the analogous term involving the nuclear angular momentum being suppressed by the large nuclear mass. Hund's case a) was recovered by  adiabatically increasing $H_{\rm pert}$. A simple analytic formula for the energy $W$ in the doublet case ($s={1}/{2}$) was obtained as a result, whose validity extended throughout the intermediate region from a) to b), thus providing an adiabatic correlation of energy levels corresponding to case a) with those of case b), as predicted by Hund \cite{hund2a} \cite{hund2b} \cite{hund2c} (whose ansatz was thus proved true). The same mathematical approach was adopted to give a theoretical explanation of the $\sigma $-type doubling phenomenon, predicted by Hund and Hulthen \cite{hult1}, this hyper-doubling of the multiplet components being obtained by removing the energy degeneracy of the $+ \sigma$ and $-\sigma$ states in a stationary molecule by means of the molecular rotation. Such an effect was, thus, a higher-order one, and then observable with a minor intensity.

H.A. Kramers, instead, studied the interaction of the electron ``spin vector with the magnetic field originating in the rotating molecule'' \cite{kramers} in order to give a theoretical explanation of the splitting observed in the rotational levels of the $^{3}S$-normal state of O$_{2}$ by Mulliken \cite{mulliken4}. By using, again, group-theoretic methods, he computed the influence of the electron spin on the stationary states of a diatomic molecule in the $S$-state by introducing a perturbation term proportional to the projection of the angular
momentum of the molecule onto the rotating spin vector ($H_{\rm pert} \propto {\bvec L}\cdot {\bvec S}$):\begin{quote}
This interaction originates in the fact that when we take the rotation into account the quantities appearing in the first terms of the general perturbation potential experience small changes proportional to the velocity of the nuclei. As a result, their contribution to the matrix elements does not vanish \cite{kramers}.
\end{quote}

\section{Quantum molecular computations and the Born-Oppenheimer approximation}
\label{bosect}

\subsection{Burrau's calculations for H$_2^+$}

The first attempt to solve a  ``molecular'' Schr\"odinger equation, describing the motion of a single electron   in the field of two fixed centers,
\begin{equation}
\la \psi +\frac{8\pi ^{2}m}{h^{2}} \left( W+\frac{e^{2}}{r_{a}}+\frac{e^{2}}{r_{b}}\right) \psi =0,
\label{d1}
\end{equation}
was carried out by Burrau in 1927 for the hydrogen molecular ion H$_2^+$  \cite{burrau}. He numerically integrated the Schr\"odinger eigenvalue problem (with separation of variables) by keeping the nuclei at a fixed distance $R$, to be considered as a parameter of the problem, and obtained the lowest energy level of the electronic motion. Then, by adding the energy of the Coulomb repulsion of the fixed nuclei, ${e^{2}}/{R}$, and upon neglecting vibrations and rotations, the variation of the total energy of the molecule as a function of the nuclear distance was determined. The minimum of the curve found in this way was interpreted as corresponding to the {\it equilibrium} separation between the nuclei and to the energy of the molecule in that electronic state. The procedure then required -- according to Born and Heisenberg \cite{bornhei} -- to relax the assumption of fixed nuclei, so that the curve mentioned was considered as the ``force law'' (potential energy) governing the rotational and vibrational motions of the molecule.

Two limiting situations were discussed. In the first case, for large internuclear separations, the system  reduced to a free hydrogen atom plus a proton, so that the corresponding electronic energy was mainly given by the Coulomb interaction between the proton and the electronic charge of the H atom. Without considering the Stark effect induced by the proton, the only contribution to this energy is due to the nuclear repulsion and the total energy of the molecule takes the value typical of the hydrogen atom, that is one Rydberg, for any value of the nuclear separation $R$. The proton, however, does induce a polarization of the H atom, and thus the energy of the proton-electron interaction becomes greater than that of the proton-proton one.

In the second case of a vanishing nuclear separation, instead, the energy of the molecule coincides with that of the ground state of the ionized helium atom.

Nevertheless, Burrau also supplied numerical values of the equilibrium separation ($R_{\rm eq}\simeq 2a_{0}$, $a_0={h^{2}}/{4\pi ^{2}me^{2}}$ being the Bohr radius), the electronic energy ($W\simeq 1.204 \, {\rm Rh}=-16.28{\rm \, eV}$) and the heat of dissociation ($D_{{\rm H}_{2}^{+}}\simeq 0.204 \, {\rm Rh}=2.76 \, {\rm eV}$) for intermediate electronic separations, and checked such values against experimental data, obtaining an encouraging agreement. He provided as well the average electron density $|\psi|^{2}$ as a function of position and depicted the corresponding contour plot, whose interpretation clearly pointed towards the concept of chemical bond, as noticed later by Pauling:
\begin{quote}
The electron is most of the time in the region between the two nuclei, and can be considered as belonging to them both, and forming a bond between them  \cite{pauling5}.
\end{quote}

Second order perturbation theory calculations on the hydrogen molecular ion was, later in 1927, performed by A. Uns\"old \cite{unsold}, who obtained numerical estimates ($R_{\rm eq}\simeq 1 a_{0}$, $W \simeq -16.31\, {\rm eV}$) in partial good agreement with the above results.

Burrau's numerical solution of the Schr\"odinger equation was also applied to the hydrogen molecule by Condon \cite{condon2}, obtaining a very good estimate of the ground term by interpolation between $0$ and $\infty$ internuclear distances.

\subsection{Variational approach}

The variational Ritz method, applied by G. Kellner to the description of the normal He atom \cite{kellner}, was employed in 1928 by Wang to solve the Schr\"odinger equation for the H$_2$ molecule \cite{wang1} in the limit of two hydrogen atoms very close together. The wavefunction for the system was assumed to be of the form
\begin{equation}
\psi = C\left[ \erm^{-Z\frac{\left( r_{1}+p_{2}\right) }{a_0}}+e^{-Z\frac{\left( r_{2}+p_{1}\right) }{a_0}}\right] ,  \label{d12}
\end{equation}
($r_{i}$ and $p_{i}$, with $i=1,2$, are the distances of the electrons $1$ and $2$ to the two nuclei, respectively), where $Z$ was treated as a variational parameter, so that for every internuclear distance a value of $Z$ exists that minimizes the energy functional and gives the best approximation of the corresponding energy value $E$. S.C. Wang indeed built up an approximate energy curve as a function of the distance, whose minimum determined his prediction (for $Z=1.166$) for the heat of dissociation ($D_{{\rm H}_{2}}=3.76 \, {\rm eV}$), the moment of inertia ($J=4.59\cdot 10^{-41} \, {\rm g} \, {\rm cm}^{2}$) and the nuclear vibrational frequency ($\nu _{0}=4900 \, {\rm cm}^{-1}$) of the hydrogen molecule in its normal state, which were in slightly better agreement with experimental data than those earlier obtained by Y. Sugiura \cite{sugiura} and Condon \cite{condon2}.

The Ritz method was successfully employed (by following the lines traced by Wang) also by B.N. Finkelstein and G.E. Horowitz \cite{horowitz} to the H$_2^+$ molecular ion, again obtaining better values ($Z=1.228$, $r_{0}=01.06 \, {\rm A}$, $W_{{\rm H}_{2}}=-15.75 \, {\rm eV}$) than those given by perturbative calculations \cite{pauling5}.

In 1929, V. Guillemin and C. Zener \cite{gz1} further improved these results by introducing a wavefunction of the form
\begin{equation}
\psi =C( \alpha ,\beta ,R) \erm^{-\frac{1}{2}R\alpha \lambda }\left( \erm^{-\frac{1}{2}R\beta \mu }+e^{\frac{1}{2}R\beta \mu }\right) ,  \label{d13}
\end{equation}
where $\lambda =\frac{r_{a}+r_{b}}{R}$ and $\mu =\frac{r_{a}-r_{b}}{R}$ are elliptical coordinates ($R$ is the internuclear distance and $r_{a}$, $r_{b}$ are the distances from nuclei $a$ and $b$ to the electron), and the variational parameters $\alpha $ and $\beta $ were chosen in such a way that the first allows for the charge density $|\psi|^2$ to concentrate about the nuclei upon decreasing $R$, while the second makes the charge density to increase between the two nuclei. In the limit of $\alpha = \beta$, they obviously recovered Finkelstein and Horowitz's results.

\subsection{Energy contribution hierarchy}

\noindent As mentioned in the previous section, the fact was evident that different order of magnitude effects contributed to the energetic terms in molecular spectra, starting from those related to the electronic motion around the nuclei (which is the largest effect), and then followed by the nuclear vibrations and rotations, respectively. This hierarchy was an obvious result of the different size of the nuclear mass with respect to that of the electrons, as envisaged already in the framework of the old quantum theory. Indeed, as early as 1924, Born and Heisenberg \cite{bornhei} realized that the energy contributions in molecules appeared as terms of increasing order in $\sqrt{{m}/{M}}$, where $m$ is the electron mass and $M$ the average nuclear mass. A drawback of such a result, however, was that it would have led to the manifestation of nuclear vibrations and rotations effects at the same order, i.e. the second one in the parameter $\sqrt{{m}/{M}}$, in contrast with known empirical evidence for smaller rotational effects. The appearance of the new quantum mechanics in 1926 paved the way to a reliable solution of this problem, which was indeed obtained one year later by Bohr and Oppenheimer \cite{bo1}.

They realized that the correct sequence of energy contributions could have been obtained systematically by means of a series expansion of the Hamiltonian of the system in terms of the quantity $\kappa =\sqrt[4]{{m}/{M}}$, the suggestion coming from the observation that the nuclear kinetic energy is proportional to the fourth power of $\kappa$ with respect to the electronic kinetic energy. Nuclear vibrations and rotations came out as second and fourth order effects, respectively, while first and third order terms disappeared (the existence of an equilibrium configuration corresponding to a minimum of the electronic energy with nuclei at rest prevented the appearance of a first order term).
\begin{quote}
The electronic energy is first to be calculated for various arrangements of the nuclei fixed in space. The stable state will then be that for which the so-calculated electronic energy added to the internuclear energy is a minimum. The nuclei will then undergo oscillations about their equilibrium positions, with the electronic and nuclear energy as the restoring potential; and the molecule as a whole will undergo rotations about axes passing through its center of mass \cite{pauling5}.
\end{quote}
The energy calculations carried out up to fourth order led to the complete decoupling of vibrational, rotational and electronic motion and allowed to obtain expressions for the eigenfunctions and the transition intensities (at the zeroth order approximation) that completely agreed with previous findings by Franck \cite{franck1} and Condon \cite{condon1} \cite{condon2}. Instead, possible couplings between the three basic types of motion had to be introduced only as effects higher than fourth order (and with the inclusion of all degeneracies of the electronic motion), which were not considered by Born and Oppenheimer.

A general quantitative theory \cite{bo1} was, then, developed, which allowed to classify rotational, vibrational and electronic terms in a molecule with $N$-atoms.

Indeed, Born and Oppenheimer assumed the potential energy $U ( x_{1}, y_{1}, z_{1}, x_{2}, y_{2},$ $z_{2}, \dots ; X_{1}, Y_{1}, Z_{1}, X_{2}, Y_{2}, Z_{2}, \dots ) \equiv U ( x, X )$ of the molecule to depend only on the relative positions of the particles, $x$ and $X$ being the electronic and nuclear coordinates, respectively, while the kinetic energy of electrons and nuclei was written as $T_{E}=-\frac{h^{2}}{8\pi ^{2}m} \sum_{x}\sum_{k}\frac{\partial ^{2}}{\partial x_{k}^{2}}$ (where the sum $\sum_{x}$ runs over all terms obtained through cyclic permutation of $x$, $y$, $z$) and $T_{K}=-\kappa ^{4}\frac{h^{2}}{8\pi ^{2}m}\sum_{X}\sum_{\ell}\mu_{\ell}\frac{\partial ^{2}}{\partial X_{\ell}^{2}}$, respectively. The total energy operator was then written as $H=H_{0}+\kappa ^{4}H_{1}$, with $T_{E}+U=H_{0}\left( x,\frac{\partial }{\partial x};X\right) $ and $T_{K}=\kappa ^{4}H_{1}\left( \frac{\partial }{\partial X} \right) $. By introducing the $3N-6$ functions $\xi _{i}=\xi_{i}\left( X\right) $ fixing the relative positions of the nuclei with respect to each other, and the $6$ functions $\vartheta _{i}\left( X\right) $
giving the orientation in space of the nuclear framework,
%with the corresponding transformation equations,
$H_{1}$ was split into three contributions, each of them with a characteristic behavior: $H_{1}=H_{\xi \xi }+H_{\xi \vartheta }+H_{\vartheta \vartheta }$. Born and Oppenheimer thus showed that the solution of the relevant eigenvalue problem,
\begin{equation}
\left( H+\kappa ^{4}H_{1}-W\right) \psi =0 \, ,  \label{p1}
\end{equation}
corresponding to a stable molecule, could be obtained through a power series in $\kappa $, and, by neglecting possible resonance degeneracy ({\it \`a la} Heisenberg), they succeeded in finding a general expression for the energy of the system up to fourth order ($n$, $s$ and $r$ are the principal, vibrational and rotational quantum numbers, respectively):
\begin{equation}
W_{nsr}=V_{n}^{0}+\kappa ^{2} \, W_{ns}^{\left( 2\right) }+\kappa^{4} \, W_{nsr}^{\left( 4\right) }+ \dots \, .  \label{p2}
\end{equation}
Here $V_{n}^{0}$ is the minimum value of the electronic energy, i.e. a characteristics of the given molecule at rest, $W_{ns}^{\left( 2\right) }$ is the energy associated to nuclear vibrations and $W_{nsr}^{\left(
4\right) }$ that of rotations. The crucial result was that a decoupling of electronic, vibrational and rotational motions occurs up to $\kappa ^{4}$ order, while higher powers of $\kappa$ should be considered in order to take into account also their coupling.

\subsection{Born-Oppenheimer approximation for diatomic molecules}

In the simplest case of a diatomic molecule, and disregarding the fine structure of the spectroscopic bands, that is by neglecting the degeneracy due to the identity of the electrons (and, in case, also that of the nuclei) and that corresponding to the self-rotation around the axis connecting the nuclei (they restricted to the case of a zero angular momentum around the molecular axis), Born and Oppenheimer obtained even more detailed results. Indeed, at the second order, the vibrational contribution in Eq. (\ref{p2}) was shown to be:
\begin{equation}
\kappa ^{2}W_{ns}^{\left( 2\right) }=\left( s+\frac{1}{2}\right) h\nu _{0} \, ,
\label{p3}
\end{equation}
where $\nu _{0}=\frac{1}{4\pi }\sqrt{\left( \frac{1}{M_{1}}+\frac{1}{M_{2}}\right) V_{n}^{^{\prime \prime }}}$ is the oscillator frequency depending on the second derivative of $V,$ while, at the fourth order, the rotational energy was obtained as a generalization of the Kramers and Pauli ansatz \cite{kep1} (who described a molecule as a rigid top):
\begin{equation}
\kappa ^{4}W_{nsr}^{\left( 4\right) }=\frac{h^{2}}{8\pi ^{2}J}g_{ns}\left(
r\right) .  \label{p4}
\end{equation}
Here $g_{ns}(r)$ is a numerical function of the rotational quantum number $r$, and $J=\frac{M_{1}M_{2}}{M_{1}+M_{2}}\xi ^{2}$ is the moment of inertia of the two nuclei in the equilibrium position.

As a result, the electronic energy of diatomic molecules was shown to be approximately independent of the rotation and vibration velocity of the nuclei, while depending only on their instantaneous relative separation. Then, within the Born-Oppenheimer picture, this allows one to neglect the motion of the nucleus in a first approximation in order to evaluate the electronic energy of the molecule, and the determination of the rotational states requires only the knowledge of the equilibrium distance between the nuclei.

Conversely, the determination of the vibrational states soon appeared much more involved, as deduced also by the failure of a previous ansatz by Fues \cite{fues}, who generalized an early proposal by A. Kratzer \cite{kratzer1} and assumed, for the corresponding potential energy $E(R)$ of a diatomic molecule, a series expansion in the internuclear distance $R$:
\begin{equation}
E(R) = \left\{
\begin{array}{l}
\displaystyle \frac{a}{R}+\frac{b}{R^{2}}+c \left( R-R_{0} \right) ^{3}+ \dots \, , \\ \\
\displaystyle b^{^{\prime }}\left( R-R_{0}\right) ^{2}+c^{^{\prime}}\left( R-R_{0}\right) ^{3}+ \dots \, .
\end{array} \right.
\end{equation}
The energy levels had, thus, to be computed by perturbation theory, so adding a further approximation in the calculation, and the series obtained for $E(R)$ did not converge for large values of $R$. Indeed, as noted later by E.A. Hylleraas \cite{hylleraas1}, Fues' ansatz was in general not very useful due to the fact that the energy potential goes as $1/R$ at the infinity, so that
\begin{quote}
the number of vibrational states becomes infinite, and the ansatz is only useful for polar molecules \cite{hylleraas1}.
\end{quote}
The choice of $E(R)$ was indeed crucial, and the following requirements \cite{morse} had to be fulfilled in order to find the correct expression: a) an asymptotically finite value as $R\rightarrow \infty $, b) the existence of only one minimum point at $R=R_{0}$, c) a divergence (or a very large value) for $R=0$, d) a finite polynomial form for the allowed energy levels.

Some years later, in 1929, P.M. Morse \cite{morse} filled this gap by proposing a quite simple form for the potential energy able to produce the ``typical'' properties of the vibrational states of the homopolar molecules,
\begin{equation}
E(R) =D \, \erm^{-2a\left( R-R_{0}\right) }-2D \, \erm^{-a\left(
R-R_{0}\right) };  \label{p5}
\end{equation}
that allowed him to solve exactly the relevant Schr\"odinger equation and thus to find the following expression for the vibrational energy levels:
\begin{equation}
W(s) =-D+h\omega _{0}\left( s+\frac{1}{2}\right) -\frac{h^{2}\omega _{0}^{2}}{4D}\left( s+\frac{1}{2}\right) ^{2}.  \label{p6}
\end{equation}
Here $\omega _{0}$ is the frequency of small vibrations and $s$ takes on all integer values from zero to ${( k-1) }/{2}$, with $k={4\pi \left( 2\mu D\right) ^{1/2}}/{ah}$ and $\mu ={M_{1}M_{2}}/(M_{1}+M_{2})$.

\

\begin{quote}
It is particularly important that the number of terms is finite \cite{hylleraas1}.
\end{quote}

\

\noindent The above energy levels were found to be in good agreement with the experimental data coming from a number of different molecules, up to quite high values of $s$, while the changes introduced for taking into account also the molecular rotation coincided in first approximation with previous results by Kratzer
\cite{kratzer1}.

\subsection{Wilson's technical analysis}

Within the zeroth order Born approximation, A.H. Wilson \cite{wilson1} carried out a detailed study of the differential equation (\ref{d1}) (in elliptical coordinates) defining the two-center problem, by focusing on its possible analytical solutions. He realized that, in general, there were no solutions bounded everywhere, and checked this conclusion against previous attempts \cite{burrau} \cite{hund2b}.

Burrau's original method \cite{burrau} was to cast that equation into a Riccati form and then solve numerically by expanding it around the irregular singularity at infinity, and Wilson noted that the expansion so obtained was divergent, though asymptotic, and no justification was provided by Burrau on its use.

On the other hand, Hund's work \cite{hund2b} was interpreted as if he would have solved a one-dimensional problem to be later generalized to separable systems, but this is certainly not the case for the equations describing a two-center problem, as noted by Wilson. However, it was recognized that Hund only used his results in order to prove that no energy term is lost when the transition he envisaged from infinitely distant nuclei to coinciding ones is performed, and such result was indeed proved by Wilson by using an asymptotic expansion of the Schr\"odinger equation.

Uns\"old, instead, followed a more standard perturbative method (hydrogen atom perturbed by a proton), but convergence problems arise in higher order approximations, since -- Wilson noted -- second order results are about twice (and with an opposite sign) those of first order (and Uns\"old did not evaluate third order terms for comparison).

From such an analysis, Wilson deduced pragmatically that the agreement of Burrau's results with experiments  testifies for the appropriateness of the method they were obtained, so that (following Burrau) he decided to relax the assumption of a solution $\psi$ that is finite throughout the whole space, but still vanishing at infinity in order to retain its physical meaning. As a result, he found solutions which become logarithmically infinite along the nuclear axis and bounded elsewhere, in full analogy with the theory of relativistic hydrogen atom; according to Pauling:
\begin{quote}
These solutions would not be considered eigenfunctions if the usual definition is retained; but would be in the case the restriction that the eigenfunction be bounded everywhere were replaced by the restriction that it be quadratically integrable \cite{pauling5}.
\end{quote}
Wilson then proved that, within such assumptions, the properties of the hydrogen molecular ion in its normal state were just those approximately given by Burrau \cite{burrau}.

\section{Method of molecular orbitals}
\label{molorb}

\noindent The relationship between the spectral terms of the component atoms and those of the molecule formed was the focus of early theoretical interpretations of spectroscopic data, with or without the proper aid of group theory. As we have seen in Sect.\ref{spectrosect}, the result of this search was a classification of all the possible molecular terms, but without any explanation about their stability. The first step towards the understanding of such a point was performed already by Hund \cite{hund1}, who, by extending previous ideas \cite{hund2a}, assigned to each electron within a molecule a definite and unique series of quantum numbers, coming out from the fact that each electron has a give angular momentum quantized along the molecular axis and from the same atomic quantum numbers which it would have if the component atoms were pushed together to form a ``united atom.''

\subsection{Mulliken's ``promoted electrons''}

Just along the lines followed by Hund, in 1928 Mulliken provided a general spectroscopic interpretation of the band spectrum \cite{mulliken1} aimed at assigning appropriate ``molecular" quantum numbers to the electrons in (non rotating) diatomic molecules.
\begin{quote}
Hund's work enables us to understand how a continuous transition can exist between ionic and atomic binding. Briefly, the molecule may be said to be latent in the separated atoms; in a certain sense, the molecular quantum numbers already exist before the atoms come together, but take on practical importance, at the expense of the atomic quantum numbers, only on the approach of the atoms to molecular distances. This of course does not exclude the possibility that in some cases a quantum jump in the usual sense may be needed to reach the most stable state of the molecule \cite{mulliken1}.
\end{quote}
The Pauli principle was recognized to hold in molecules as well, and its application allowed to classify the possible molecular states corresponding to a given electron configuration:
\begin{quote}
This is done by Pauli's method of imagining the atom in an external magnetic field so strong that all couplings between electrons are broken down, so that each electron can be given four quantum numbers, $n$, $\ell$, $m_{\ell}$ and $m_{s}$. In a molecule, this breaking down is partly accomplished by the intramolecular electric field \cite{mulliken1}.
\end{quote}

In a sense, Mulliken's idea was to study the group theory properties of spectral terms, but {\it without} the use of group theory.

The relative energies of the different electronic levels, which can be obtained -- in the limiting case of zero
internuclear distance -- when molecular states collapse on atomic states, were deduced theoretically by Hund  \cite{hund1}, but, in the more involved case of intermediate internuclear distances, no simple limit exists, and the energy levels were obtained by interpolation and then fitted to the experimental data, in order to produce a smooth transition between the two extreme limiting cases (zero or infinity internuclear distance). Both Hund and Mulliken realized as well that the table of energy levels had to be supplemented by certain empirical rules, but while Hund \cite{hund1} tried to limit the intersections of the possible transitions from a given state of the separated atoms to the corresponding states of the united atom, Mulliken \cite{mulliken1} adopted the hypothesis of a constant number of $\sigma $, $\pi $, ... electrons during such transitions.

In order to improve the results obtained, Mulliken tried to determine and classify the electronic states of the atomic products (i.e. one excited atom or ion) resulting from the dissociation of unexcited molecules when increasing the vibrational quantum number of the corresponding states. Such a generalization was aimed at providing an explanation of some experimental results by Herzberg on the N$_{2}^{+}$ molecule
\cite{herzberg2}, that is N$_{2}^{+}$ in its ground state got split into a normal N atom and an excited N ion. Similar results were found also for CN, CO$^{+}$, BO and SiN. A further generalization dealt with diatomic
hydride molecules of the form $M$H \cite{mulliken2}, whose electronic states were classified by Mulliken according to the effects produced by the H on the $M$ atom. The main effects he identified were as follows: 1) couplings between $\ell_{r}$ vectors of the outer electrons of $M$ are completely broken down by the field of the H nucleus and the usual $\ell_{r}$ selection rules were superseded; 2) the uncoupled vectors $\ell_{r}$ are separately quantized with respect to the electric axes, giving rise to component quantum numbers $i_{\ell_{r}}$; 3) the electron of the H atom ($i_{\ell_{r}}=0$) is ``promoted" and sometimes forms a closed shell with one of the $M$ electrons, while the H nucleus is on the outside edge of the $M$ electron cloud: as a consequence the hydrides are strongly polar molecules; 4) the original couplings of $s_{r}$ vectors are often broken down because of the presence of the H electron spin, and the original multiplicity gets modified by one unit.

In Mulliken's own words, the basic assumption of his theory was that
\begin{quote}
molecular stability is primarily a matter of promotion energy, rather than of valence bonds in the sense of Lewis or London \cite{mulliken2}.
\end{quote}

\subsection{Lennard-Jones and the molecular aufbau}

Mulliken looked at a molecule as at a collection of nuclei fixed in given spatial locations and surrounded by an electron cloud \cite{mulliken2} \cite{mulliken3}, and it was soon easy to take the next step and thus embed each electron into a ``molecular orbital" which spreads over the whole molecule: as the atoms approach each other, they loose their identity and  share electrons, giving rise to the chemical bond. The number of adherents to this novel view indeed grew up for some time, producing key contributions as well \cite{herzberg1} \cite{huckel} \cite{LJ}.

In 1929, Lennard-Jones \cite{LJ} further developed Mulliken's ideas, and introduced a clear distinction between atomic levels and molecular levels within a molecule: molecular levels give rise to binding while atomic ones produce repulsion due to an asymmetry of the corresponding eigenfunctions (the last concept being borrowed by Heitler-London; see below). He then set up an aufbau method for molecules analogous to that for atoms:
\begin{quote}
We suppose a molecule built up in the following way. We add one charge at a time to each nucleus and then, supposing the nuclei held fixed, add to the system two electrons successively. The system is then allowed to take up its equilibrium value adiabatically. Next, we add the components of the angular momenta about the nuclear axis [...]. Then, as in atoms, we add the electron spins to determine the multiplicity \cite{LJ}.
\end{quote}
The effect of core electrons was supposed to produce only an imperfect screening of the charge of the nucleus, depending on the configuration of the outer electron, and the effective charge had to be adjusted in order to give rise to the observed energy level for the atomic states. Vibrational and rotational effects were, instead, assumed to be small perturbations on the electronic configuration, and thus negligible. Of course, that didn't apply to the light molecule He$_{2}$, so that Lennard-Jones' work was mainly devoted to describe the ground levels or excited states of heavier molecules. In particular, he explicitly built up the electronic structures of the series of diatomic molecules and molecular ions from Li$_{2}$ to F$_{2}$ (but also applied his method to H and He molecular compounds).

His analysis was supplemented by the following rules, devised to assign properly each electron to a lowest energy cell:
\begin{quote}
I. Whenever two electrons are placed in a molecular cell, the corresponding cell of each atom must be left open.
\\
II. Whenever two electrons are allotted to the same atomic cell, the corresponding molecular cell must be excluded. [...]
\\
III. Whenever two electrons are in a molecular state, they must separate on dissociation so that one goes over to one atom and one to the other, the electron spins being opposite; and when several pairs of electrons are in
molecular states, they must separate so that one electron from each cell goes over to each atom, and the spins of those electrons which go over to the same atom must all point in the same direction. [...]
\\
IV. Whenever one electron appears in one molecular cell and one in another, a resonance between molecular states must be considered, and that state will have the lowest energy which corresponds to the greatest multiplicity \cite{LJ}.
\end{quote} %An interesting conclusion was the existence of two kinds of  ``unsaturated valencies'' able to produce molecular binding, the first one being due to an incomplete group of electron spins while the second one due to an incomplete group of angular momentum spins.

\subsection{Wavefunctions}

Hund's and Mulliken's original ideas about molecular orbitals were carried forward intensely by them and other authors in the years following their initial papers, with the intention ``to describe and understand molecules in terms of one-electron orbital wavefunctions of distinctly molecular character'' \cite{mulli32}. However, as apparent from the above, their works were mainly qualitative in nature, aimed to deal with spectroscopic data in the ``style'' of the old quantum theory,\footnote{Particularly illuminating is what Mulliken wrote in one of his reviews:  ``For the sake of simplicity and  `Anschaulichkeit,' the treatment in Parts I-II is in terms of the {\it old quantum theory} and repeatedly involves the use of models which, according to the new quantum theory, must not be taken too literally. So far as possible, however, the most essential new result of the new quantum theory -- especially energy relations -- are stated in the text, although their rigorous derivation is not given.''} and the implementation of the molecular orbital concept into a mathematical formulation suitable for application to quantum mechanics took some time to be performed. However, the method of molecular orbitals originally devised by Hund, Mulliken and Lennard-Jones, was eventually translated into a method for the writing of appropriate molecular wavefunctions (following, in a sense, the general approach -- but certainly not the spirit -- of Heitler and London \cite{heitler1}) to be used in suitable approximations to be adopted in order to solve the relevant Schr\"odinger equation.

An especially clear formulation was provided later (in 1935) by van Vleck \cite{history1}, from whom we quote:
\begin{quote}
A {\it molecular orbital} is defined as a wavefunction which is a function of the coordinates of only one electron, and which is, at least hypothetically, a solution of a dynamical problem involving only one electron. The method of molecular orbitals seeks to approximate the wavefunction of a molecule containing $n$ electrons as the product of $n$ molecular orbitals, so that
\begin{equation}
\Psi =\psi _{1}( x_{1},y_{1},z_{1}) \psi _{2}( x_{2},y_{2},z_{2}) \cdots \psi _{n}( x_{n},y_{n},z_{n}) .
\label{b1}
\end{equation}
\end{quote}
For a diatomic molecule, in particular, this becomes:
\begin{equation}
\Psi =\left[ a \, \psi _{A}( 1) +b \, \psi _{B}( 1) \right]   \left[ a \, \psi _{A} ( 2) +b \, \psi _{B}( 2) \right] ,
\label{b2}
\end{equation}
where $\psi _{A}$ and $\psi _{B}$ are {\it atomic orbitals} for atoms $A$ and $B$,
respectively, since, according to the clear definition of van Vleck, ``by atomic orbital is meant one-electron wavefunction for an electron moving in a field of only one atom" \cite{history1}. By rewriting Eq. (\ref{b2}) as follows:
\begin{equation}
\Psi =a^{2}\psi _{A}( 1) \psi _{A}( 2) +b^{2}\psi_{B}( 1) \psi _{B}( 2) +ab\left[ \psi _{A}(1) \psi _{B}( 2) +\psi _{A}( 2) \psi _{B}(1) \right] ,  \label{b3}
\end{equation}
it is evident that terms such as $\psi _{A}( 1) \psi _{A}( 2) $ and $\psi _{B}( 1) \psi _{B}( 2) $ imply that electrons $1$ and $2$ are both on the atom $A$ or on the atom $B$, so that they represent ionic terms, while the remaining terms $\psi _{A}( 1) \psi_{B}( 2) $ and $\psi _{A}( 2) \psi _{B}( 1)$ are characterized by one electron on each atom, and thus feature polar terms.

These parameterizations should be regarded as unperturbed wavefunctions to be used as starting points in perturbative calculations of the Schr\"odinger equation, but it turned out \cite{history1} that such a method of molecular orbitals revealed to be much more amenable to a qualitative discussion than to quantitative calculations, for which the competing method introduced by Heitler and London as early as 1927 was largely adopted.

\section{The Heitler-London theory}
\label{hlsect}

\begin{quote}
The mathematical problem of a chemical reaction seems to be this: to investigate whether there are stable solutions of the Schr\"odinger wave equation corresponding to the interaction between two (or more) atoms, using only the wavefunctions which have the type of symmetry compatible with Pauli's exclusion principle. [...] Recent work of London and Heitler seems to indicate that the systematization of chemical compounds is closely related to the group theory of mathematicians \cite{vanvleck}.
\end{quote}
The Heitler-London theory \cite{heitler1} \cite{heitler2} \cite{london1} \cite{london2} of homopolar molecules, indeed, was primarily based on symmetry properties of the wavefunctions as required by the Pauli exclusion principle, but the key novel idea was to extend Heisenberg's idea of resonance \cite{Hei26} \cite{hei1} by introducing purely quantum-mechanical {\it exchange forces}, as opposed to polarization forces. The rationale of their work was just the quest for a theoretical understanding of the interaction between neutral atoms (of which homopolar molecules are composed of), which led them to take into account the possibility of a non-polar binding.

Heitler and London investigated this issue by making explicit reference to the simplest examples of  H$_2$ and He$_2$, and in the following we will follow closely the main steps of their work.

\subsection{H$_2$ molecule}

The first problem they considered was the determination of the change in energy as experienced by two neutral hydrogen atoms in their ground states, when approaching each other up to a fixed distance $R$ (taken as the distance between the two nuclei). The corresponding Schr\"odinger equation for the system of two nuclei $a$ and $b$ and two electrons $1$ and $2$,
\begin{equation}
\left( \la_{\!\!\!\!\!\! 1} \,\, + \la_{\!\!\!\!\!\! 2} \,\, \right) \chi +\frac{8\pi ^{2}m}{h^{2}}%
\left[ E-\left( \frac{\varepsilon ^{2}}{R}+\frac{\varepsilon ^{2}}{r_{12}}-%
\frac{\varepsilon ^{2}}{r_{a1}}-\frac{\varepsilon ^{2}}{r_{a2}}-\frac{%
\varepsilon ^{2}}{r_{b1}}-\frac{\varepsilon ^{2}}{r_{b2}}\right) \right]
\chi =0 \, ,  \label{p7}
\end{equation}
had to be solved, and, to this end, Heitler and London adopted a simple perturbation theory to be applied to unperturbed wavefunctions built up from the well-known eigenfunctions $\psi _{i}$, $\varphi _{i}$ of the hydrogen atom for the $i$-electron ($i=1,2$) present on nucleus $a$, $b$.\footnote{That is, $\psi _{i}=\frac{1}{\sqrt{\pi } \, a_0^{3/2}} \, \erm^{-\frac{r_{ai}}{a_{0}}}$ and $\varphi _{i}=\frac{1}{\sqrt{\pi } \, a_0^{3/2}} \, \erm^{-\frac{r_{bi}}{a_{0}}}$.} As unperturbed eigenfunctions they chose those characterized by one electron on the first nucleus and the other electron on the other nucleus, but since a two-fold degeneracy is present, the two following linear combinations  were considered: $\alpha = a \, \psi _{1} \varphi _{2} + b \, \psi _{2} \varphi _{1}$ and $\beta = c \, \psi _{1} \varphi _{2} + d \, \psi_{2}\varphi _{1}$. The numerical coefficients were determined by means of the usual normalization and orthogonality conditions, the result being:
\begin{equation}
\left\{
\begin{array}{c}
\dis \alpha = \frac{1}{\sqrt{2+2S}}\left( \psi _{1}\varphi _{2}+\psi _{2}\varphi_{1}\right) , \\ \\
\dis \beta =\frac{1}{\sqrt{2+2S}}\left( \psi _{1}\varphi _{2}-\psi _{2}\varphi_{1}\right) ,
\end{array} \right.  \label{p9}
\end{equation}
with normalization factors expressed in terms of the integral $S = \int \! \psi _{1}\varphi _{1}\psi _{2}\varphi _{2} \, \drm \tau _{1}\drm \tau _{2}$. From the expression above, it is easily recognizable that the two combinations $\alpha$ and $\beta$ are symmetric and antisymmetric in the exchange of the two electrons $1$ and $2$ (or the two nuclei), respectively. The perturbation theory was, then, applied, and the two perturbed energies corresponding to the two states above were obtained:
\begin{equation}
\left\{ \begin{array}{c}
\dis E_{\alpha }=E_{11}-\frac{E_{11}S-E_{12}}{1+S} \, , \\ \\
\dis E_{\beta }=E_{11}+\frac{E_{11}S-E_{12}}{1-S} \, ,
\end{array} \right.   \label{p11}
\end{equation}
where
\begin{eqnarray}
E_{11} &=& \int \left[ \left( \frac{e ^{2}}{r_{12}}+\frac{ e ^{2}}{R}\right) \frac{\psi _{1}^{2}\varphi _{2}^{2}+\psi_{2}^{2}\varphi _{1}^{2}}{2}-\left( \frac{e ^{2}}{r_{a1}}+\frac{ e ^{2}}{r_{b2}}\right) \frac{\psi _{2}^{2}\varphi _{1}^{2}}{2}  \right. \nonumber \\
& & \left. \qquad -\left( \frac{e ^{2}}{r_{a2}}+\frac{e ^{2}}{r_{b1}} \right) \frac{\psi _{1}^{2}\varphi _{2}^{2}}{2}\right] \drm \tau _{1} \drm \tau _{2}, \label{p12} \\
E_{12} &=&\int \left( \frac{2e ^{2}}{r_{12}}+\frac{2e
^{2}}{R}-\frac{e ^{2}}{r_{a1}}-\frac{e ^{2}}{r_{a2}}-%
\frac{e ^{2}}{r_{b1}}-\frac{e ^{2}}{r_{b2}}\right) \frac{%
\psi _{1}\varphi _{1}\psi _{2}\varphi _{2}}{2} \, \drm \tau _{1} \drm \tau _{2}.
\label{p13}
\end{eqnarray}
Here, the key result was the lifting of the initial degeneracy as shown in Eqs. (\ref{p11}), and since the eigenfunction of atom $a$ does not vanish in the spatial position occupied by atom $b$ and viceversa, Heitler and London deduced that a finite probability exists for the electron of atom $a$ to belong to $b$.  A resonance phenomenon {\it \`a la} Heisenberg \cite{Hei26} \cite{hei1} took place, $[h ( E_{\beta }-E_{\alpha})]^{-1}$ being the frequency of the average exchange between electrons.
\begin{quote}
While in classical mechanics it is possible to label the electrons (we put each electron in a sufficiently steep potential well and do not allow energy addition), something similar is impossible in quantum mechanics: when at one moment in time one is certain to know one electron in the potential well, one can never be certain that in the next moment it does not exchange with another \cite{heitler1}.
\end{quote}
Or, in different words:
\begin{quote}
In the hydrogen molecule the average electronic charge distribution is symmetrical with respect to the two nuclei, and the two electrons are continually exchanging places, so that it is impossible to say which electron belongs with which nucleus \cite{vanvleck}.
\end{quote}
Heitler and London \cite{heitler1} realized that the solution $\alpha $ doesn't have any node while the antisymmetric one $\beta $ always has one node, so that $E_{\beta }>E_{\alpha }$. However, they also proceeded to give a quantitative estimate of such energies, starting from the calculation of $E_{11}$ and $E_{12}$, but, while $E_{11}$ was clearly interpreted as due to the pure Coulomb interaction of the charge distributions present in the molecule, the meaning of $E_{12}$ was not so transparent. And, while they carried out the full calculation for $E_{11}$,\footnote{Obtaining
$
E_{11}=\frac{e ^{2}}{a_{0}} \, \erm^{-\frac{2R}{a_{0}}}\left( \frac{a_{0}%
}{R}+\frac{5}{8}-\frac{3}{4}\frac{R}{a_{0}}-\frac{1}{6}\frac{R^{2}}{a_{0}^{2}%
}\right) . %\label{p14}
$}
only an upper limit for the integral $\int \frac{\psi _{1}\varphi_{1}\psi _{2}\varphi _{2}}{r_{12}} \drm \tau _{1} \drm \tau _{2}$ appearing in the expression for $E_{12}$ was obtained, leading only to an estimate of this energy term. The exact evaluation of $E_{12}$ was instead carried out some months later by Sugiura \cite{sugiura},\footnote{Who found
$%\begin{equation}
E_{12}=\frac{e ^{2}}{a_{0}}\left[ S \, \frac{a_{0}}{R}-\erm^{-\frac{2R}{a_{0}}}\left( \frac{11}{8}+\frac{103}{20} \frac{R}{a_{0}} +\frac{49}{15} \frac{R^2}{a_{0}^2}+\frac{11}{15}\frac{R^3}{a_{0}^3}\right) +\frac{5}{6R}\left( S\left( C+\log R\right)  + \right. \right. $
$\left. \left. S^{\prime } {\rm Ei} \left( -4R\right) -2\sqrt{S S^{\prime }} {\rm Ei}\left( -2R\right) \right) \right] ,  %\label{p15}
$ %\end{equation}
with $S=\left( 1+ \frac{R}{a_{0}}+ \frac{1}{3} \frac{R^2}{a_{0}^2}\right) ^{2}\erm^{-\frac{2R}{a_{0}}}$, $S^{\prime }=\left( 1- \frac{R}{a_{0}}+\frac{1}{3}\frac{R^2}{a_{0}^2}\right) ^{2} \cdot $ $\erm^{-\frac{2R}{a_{0}}}$, $C=0,57722$ being the Euler constant and ${\rm Ei} \left( -x\right)$ the  exponential integral function, respectively.}
%=\left\{ \begin{array}{ll} C+\frac{1}{4}\log x^{4}-x+\frac{1}{2}\frac{x^{2}}{2!}-\frac{1}{3}\frac{x^{3}}{3!}+ \dots , & x\leq 17 , \\ & \\ \frac{e^{-x}}{-x}\left( 1-\frac{1!}{x}+\frac{2!}{x^{2}}-\frac{3!}{x^{3}} + \dots \right), & x>17. \end{array} \right. .  \label{p16} \end{equation}}
thus allowing also the explicit evaluation of the moment of inertia and the oscillation frequency (and then the dissociation energy) of the hydrogen molecule, and, in the appropriate limit, also the ionization energy of the helium atom.

\begin{figure}
\begin{center}
\includegraphics[width=8cm]{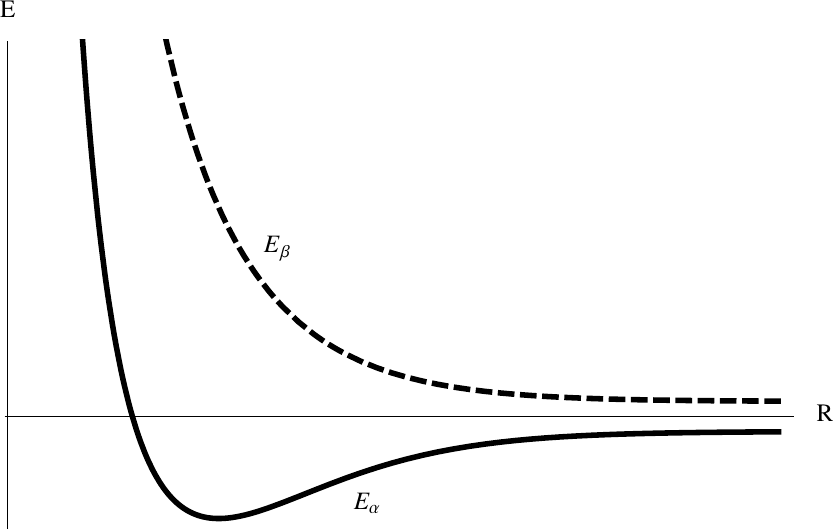}
\caption{Heitler-London energies for the symmetric and antisymmetric configurations of the hydrogen molecule, as a function of the internuclear distance $R$.}
\label{fig1}
\end{center}
\end{figure}

However (and correctly), here the focus was not on  numerical predictions for the hydrogen molecule, but rather on the physical interpretation of the two solutions $\alpha$, $\beta$ obtained so far, corresponding to the interaction energy functions reported in Fig. \ref{fig1}. Heitler and London interpreted the antisymmetric solution with energy $E_{\beta }$ as representing the van der Waals repulsion between the two hydrogen atoms (``elastic reflection''), while that corresponding to the attractive potential $E_\alpha$ was identified with the formation of a stable homopolar molecule, the minimum of $E_{\alpha }$ representing the equilibrium configuration.

Heitler and London thus concluded that ``the non-polar attraction is a characteristic quantum-mechanical effect'' \cite{heitler1}, driven just by their exchange interaction.

\subsection{He$_2$ molecule}

The interaction between two helium atoms was dealt with as well by the same authors. Here the basic units are the eigenfunctions $\psi $ and $\varphi$ for the two nuclei $a$ and $b$, each one corresponding to two electrons, six combinations of their products being, in general, possible: $\psi _{12}\varphi _{34}$, $\psi _{34}\varphi _{12}$, $\psi _{13}\varphi _{42}$, $ \psi _{42}\varphi _{13}$, $\psi _{14}\varphi _{23}$, $\psi _{23}\varphi_{14} $. Heitler and London, however, realized that two of them, $\psi _{13}\varphi _{42}$ and $\psi _{42}\varphi _{13}$, were forbidden by the Pauli principle, so that they introduced only the following linear combinations:
\begin{equation}
\begin{array}{cc}
\Psi _{1}=\psi _{12}\varphi _{34}+\psi _{34}\varphi _{12}, & \Phi _{1}=\psi
_{12}\varphi _{34}-\psi _{34}\varphi _{12}, \\
\Psi _{2}=\psi _{14}\varphi _{23}+\psi _{23}\varphi _{14}, & \Phi _{2}=\psi
_{14}\varphi _{23}-\psi _{23}\varphi _{14} ,
\end{array}
\label{p17}
\end{equation}
and, within the same reasoning as for the H$_2$ molecule, the appropriate unperturbed wavefunctions resulted to be:
\begin{equation}
\begin{array}{rcl}
\dis \alpha \!\!\! &=& \!\!\! \dis \Psi _{1}+\Psi _{2}=\psi _{12}\varphi _{34}+\psi _{34}\varphi_{12}+\psi _{14}\varphi _{23}+\psi _{23}\varphi _{14}, \\ %& & \\
\dis \beta \!\!\! &=& \!\!\! \dis \Psi _{1}-\Psi _{2}=\psi _{12}\varphi _{34}+\psi _{34}\varphi_{12}-\psi _{14}\varphi _{23}-\psi _{23}\varphi _{14}, \\ %& & \\
\dis \gamma \!\!\! &=& \!\!\! \dis \Phi _{1}= \hphantom{\, \Psi _{2}=}\psi _{12}\varphi _{34}-\psi _{34}\varphi _{12}, \\ %& & \\
\dis \delta \!\!\! &=& \!\!\! \dis \Phi _{2}=\hphantom{\,\, \Psi _{2}=\psi _{12}\varphi _{34}+\psi _{34}\varphi_{12}+}\psi _{14}\varphi _{23}-\psi _{23}\varphi _{14}.
\end{array}
\label{p18}
\end{equation}
They also found that $\gamma$ and $\delta$ were two-fold degenerate in energy ($E_{\gamma}=E_{\delta }$), and then had to be discarded, thus remaining with the non degenerate wavefunctions $\alpha $ and $\beta $ ($E_{\alpha }\neq E_{\beta }$). As for the hydrogen molecule problem, also here $E_{\alpha }$ was identified with the lowest energy eigenvalue, so that the state $\alpha$ would represent (at zeroth order) the formation of a stable molecule, while $\beta$ the elastic repulsion between the two helium atoms.

Nevertheless, Heitler and London realized that $\alpha $ did {\it not} satisfy the requirements of the Pauli principle,
since two He atoms (and, in general, two noble gas atoms) cannot be distinguished with respect to their spin, contrary to what happens for hydrogen (and for all atoms with open shells). For He$_2$, the only allowed solution would correspond to $\beta$, but since this describes an unstable configuration, the molecule formation was predicted to be forbidden.

Of course, a linear combination of the solutions $\alpha $, $\beta $, $\gamma $, $\delta $ could be alternatively introduced but, likewise, it could not lead to the formation of a stable molecule because, as recognized by Heitler and London, that would correspond to four electrons in a $K$-shell, i.e. a configuration again forbidden by the Pauli principle. A solution could, then, be found only with excited helium atoms, and the conclusion was that, on the basis of the new quantum mechanics, no stable He$_{2}$ molecule could form. In Pauling's words:
\begin{quote}
It is of particular significance that the straightforward application of the quantum mechanics results in the unambiguous conclusion that two hydrogen atoms will form a molecule but that two helium atoms will not; for this distinction is characteristically chemical, and its clarification marks the genesis of the science of sub-atomic theoretical chemistry \cite{pauling5}.
\end{quote}

\subsection{Understanding the chemical bond}

London and Heitler further investigated the issue of the formation of the homopolar chemical bond in a subsequent series of papers \cite{london1} \cite{london2} \cite{heitler3}, and -- especially London -- pointed out the crucial role played by group theory. Valence numbers of homopolar molecules were interpreted  in the framework of quantum mechanics as group theory properties of the given problem, and coming out directly from the symmetry properties of the eigenfunctions of the component atoms.
\begin{quote}
The valence number is therefore given by the number of ones in the decomposition of the electronic number of an atom, which represents the symmetry character of the atomic state considered \cite{london1}.
\end{quote}
Valence was, thus, determined from the symmetries of the {\it unperturbed} atoms in the molecule, i.e.
showed up already in zeroth order eigenfunctions (and first order energies), the different modes of the homopolar binding being related to the ``lifting of the exchange degeneracy''. Higher order terms describe only effects concerning the molecule already formed, such as condensation, crystallization or van der Waals attraction.

\begin{figure}
\begin{center}
\includegraphics[width=10cm]{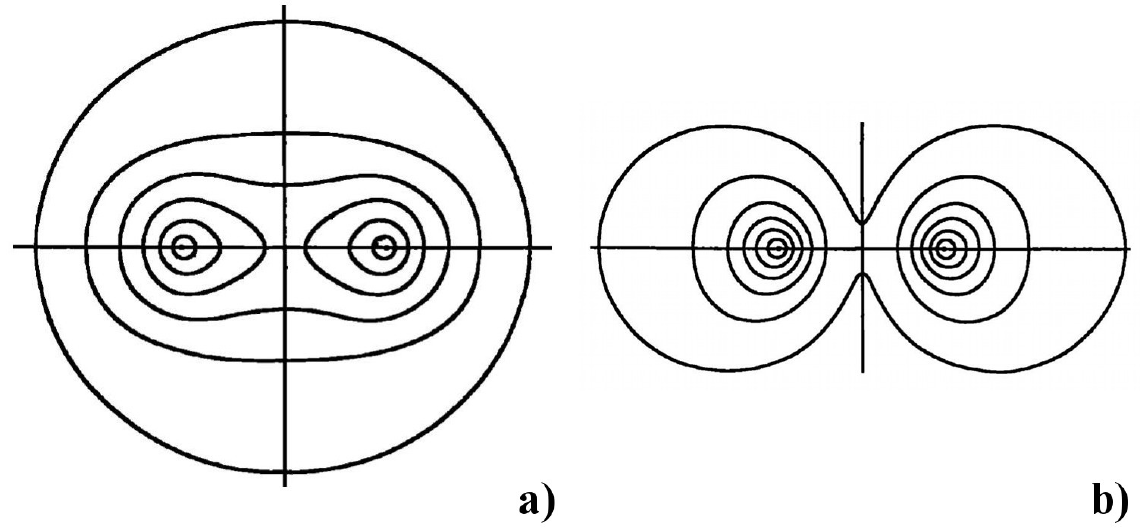}
\caption{Contour plots of the electron charge density distribution in the H$_2$ molecule. Case a): homopolar bond solution $\alpha$. Case b): elastic repulsion solution $\beta$.}
\label{fig2}
\end{center}
\end{figure}

It is a merit of London \cite{london1} to have introduced also a friendly visualization of ``the process of homopolar valence formation" which had quite a good fortune, since it helped substantially scholars to understand qualitatively the physical picture of the quantum theory of the chemical bond. It consisted merely in sketching  the contour plot of the electron charge density distribution in the molecule, calculated just with zeroth order wavefunctions. For the H$_2$ molecule, the two plots corresponding to the symmetric (elastic repulsion) and antisymmetric (homopolar bond) solutions in Eqs. (\ref{p9}) are as in Fig. \ref{fig2}.
\begin{quote}
[Fig. 2b] shows the case of two atoms that can{\it not} form a {\it bond} with each other. [...] We see that the densities are clearly pushed outward, as if they would separate if possible. If we would bring the nuclei, which are here at the same distances in the H$_2$ molecule, closer together, the strangling of the density between the atoms would increase; if the approach becomes even closer, the density here will fall to zero. [...]

In opposition to this, [Fig. 2a] shows two atoms which are in a state of {\it homopolar binding}. Here the two densities seem to draw closer and become one. With the help of these figures, we can imagine how in complicated molecules the atoms which form a valence are connected by such a bridge of $\psi \ov{\psi}$-density, while all remaining atoms stay separate \cite{london1}
\end{quote}

The (qualitative) extension to molecules more complex than H$_{2}$ and He$_{2}$ was, indeed, considered,  and the main findings were that: a) inert gases cannot exhibit valences; b) halogens may have the valences $1$, $3$, $5$, $7$ except that the valence of fluorine may only assume the value $1$; c) in the oxygen group, S, Se and Te may have valences $0$, $2$, $4$, $6$ but O only values $0$ and $2$; d) in the nitrogen group, P, As, Sb, Bi may have $1$, $3$, $5$ but N only $1$ and $3$; e) in the carbon group, C, Si, Ge may have $0$, $2$, $4$.

Instead, a generalization to atoms of the first and second column of the periodic table was provided by Pauling \cite{pauling5}. He argued that the interaction between two alkali metal atoms is similar to that between two hydrogen atoms: while the closed shells of the ions experience van der Waals-like forces typical of noble gases, the two valence electrons are shared between the two ions as in the original Heitler-London theory, and a molecule sets up in close agreement with band spectra observations. Conversely, two atoms with two valence electrons, such as Hg, interact as two helium atoms, thus giving rise only to very small attractive forces.

Again, as pointed out in 1929 by van Vleck, general results are just the manifestation of symmetries inherent to the problem at hand:
\begin{quote}
The different valences correspond to different apportionments of various values of the quantum numbers $k$, $m_{k}$ among the electrons, and the relative prevalence of the different valences depends upon the relative prevalence of the states corresponding to different values of the quantum numbers $k$, $m_{k}$ but given $n$. Some of these states may have such high energies that they are occupied only very infrequently, and so the
corresponding valences may not exist. [...]

London's work, in fact, seems to show that there is a very intimate connection between valences and the spectroscopists' classification of spectral terms \cite{vanvleck}.
\end{quote}

\

The guidance of group theory was especially required in the quantum mechanical study of non simple molecules, since the mathematical complexities of the corresponding $n$ body problem prevented the integration of the relevant Schr\"odinger equation with a suitable accuracy, as instead happened for H$_2$. After some ``metabolization'', the two competing methods of molecular orbitals and of Heitler-London were finally regarded as different approximations within which the unperturbed wavefunctions of the molecular system could be chosen. In a 1935 accurate review by van Vleck, this was explicitly recognized, along with a thorough, physical comparison of the two methods:
\begin{quote}
Because of the term $e^2/r_{12}$ in the Hamiltonian function, which represents the familiar Coulomb repulsion, two electrons dislike being close together. [...] The great failing of the method of molecular orbitals is the excessive presence of ionic terms, due to inadequate allowance for the $r_{12}$ repulsion. [...]

To avoid this difficulty of inadequate recognition of the $r_{12}$ effect, the Heitler-London method goes to the other extreme, and assumes as its defining characteristic that all ionic terms are completely wanting. [...]

The Heitler-London method is much preferable at very large distances of separation of the atoms, at least in symmetrical molecules, for then the continual transfer of electronic charge from one atom to another demanded by the ionic terms surely scarcely occurs at all. On the other hand, at small distances, the Heitler-London method probably represents excessive fear of the $r_{12}$ effect \cite{history1}.
\end{quote}

\section{Further on hydrogen and helium molecules}
\label{further}

\noindent The ideas about the chemical bond emerging from the Heitler-London theory, along with quantitative calculations carried out mainly in the Born-Oppenheimer scheme, stimulated further quantum mechanical studies on the simplest molecules, i.e. those built up with hydrogen and helium.

\subsection{H$_2^+$ ion and one-electron bond}

A perturbative approach to the solution of Schr\"odinger equation for H$_{2}^{+}$ was developed by Pauling \cite{pauling5}, who followed the Heitler-London method \cite{heitler1} and introduced similar unperturbed eigenfunctions of the form
\begin{equation}
\left\{
\begin{array}{l}
\dis \alpha =\frac{1}{\sqrt{2+2S}}\left( \psi +\varphi \right)  , \\ \\
\dis \beta =\frac{1}{\sqrt{2+2S}}\left( \psi -\varphi \right) ,
\end{array}
\right.   \label{d8}
\end{equation}
where $\psi $ and $\varphi $ are hydrogen wavefunctions corresponding to nucleus $a$ and $b$, respectively. The first order perturbation energy was found to be (including the internuclear interaction)
\begin{equation}
W^{1}=-e^{2} \, \frac{I_{1}+I_{2}}{1+S} +\frac{e^{2}}{R} \, ,  \label{d10}
\end{equation}
with $I_{1}=\int \! \frac{\psi ^{2}}{r_{A}} \drm \tau $ and $I_{2}=\int \! \frac{\psi
\varphi }{r_{A}}\drm \tau $. Numerical estimates provided by Pauling ($R_{\rm eq}\simeq 2.5a_{0}$, $W\simeq -15.30 \, {\rm eV}$, $D_{{\rm H}_{2}^{+}}\simeq 1.62 \, {\rm eV}$) were compared with Burrau's
results, thus recognizing the crucial role of quantum resonance in the formation of chemical bond:
\begin{quote}
This resonance energy leads to molecule formation only if the eigenfunction is symmetric in the two nuclei \cite{pauling5}.
\end{quote}

The idea of quantum resonance in H$_{2}^{+}$ was generalized some years later, in 1931, by Pauling himself \cite{pauling2}, by introducing the concept of {\it one-electron bond}. Indeed, resonance was introduced in order to explain the chemical bond mediated by a pair of electrons and was realized effectively by their identity, so that it should not take place in molecules with only one electron. Nevertheless, Pauling noted that, in H$_2^+$, such a phenomenon occurred alike, since the corresponding unperturbed system is degenerate in energy (the two nuclei have the same charge, and thus the same energy). This induced him to formulate the  following rule:
\begin{quote}
A stable one-electron bond can be formed only when there are two conceivable electronic states of the system with essentially the same energy, the states differing in that for one there is an unpaired electron attached to one atom, and for the other the same unpaired electron is attached to the second atom \cite{pauling2}.
\end{quote}
By analyzing the values of the dissociation energy of different compounds, he finally suggested that one-electron bonds could be present also in H$_{3}^{+}$, Li$_{2}^{+}$, boron hydrides and so on, in this way providing the basics for the understanding of the nature of the chemical bond in more complex systems.

Further investigations on H$_{2}^{+}$ by Morse and E.C.G. Stueckelberg \cite{morse1}, Lennard-Jones \cite{LJ} and Teller \cite{teller1} dealt with calculations of excited states, rather than the ground state.

Morse and Stueckelberg \cite{morse1} evaluated the electronic energies $W_{\rho}( n_{y},n_{\phi },n_{z}) $ as a function of the nuclear separation $c=2\rho $ for different values of the quantum numbers $n_{y},n_{\phi },n_{z}$, within the Born-Oppenheimer approximation \cite{bo1}. The corresponding wavefunction was separable in elliptical coordinates, and a preliminary analysis of the behavior of its nodal surfaces upon varying $\rho $ from $\infty $ to $0$ allowed them to infer some qualitative information
on the behavior of $W_{\rho }( n_{y},n_{\phi },n_{z}) $:
\begin{quote}
The number of nodes in any coordinate at a finite distance from the nucleus corresponds to the quantum number in that coordinate, and in the case considered the total number of nodes plus one equals the total quantum number $n$. The meaning of these quantum numbers in terms of a molecular model has been discussed by Mulliken \cite{morse1}.
\end{quote}
Quantitative predictions were then derived by means of two different perturbation methods originated by suitable zeroth order wavefunctions. In the first case, the wavefunction of the molecule was assumed to be that of the united atom at $\rho =0$, and the perturbation of the energy levels perturbation was evaluated upon increasing $\rho$:  although the range of validity of this calculation was restricted to small values of $\rho $, it allowed to derive a clear indication about the splitting of the different energy levels of the united atoms. Conversely, in the second case, a modification of the Heitler-London method was worked out by taking the appropriate unperturbed wavefunction to be that corresponding to infinite internuclear distance $\rho =\infty $, and assuming its validity to hold also for smaller distances: changes in the energy levels were here due  to the slight effect of the distant nucleus on the electron. Finally, the results in the intermediate region $0<\rho <\infty $ were inferred by interpolation, and the whole energy curves built up. Only the three curves corresponding to the states $1s\sigma $, $3d\sigma $ and $4f\sigma $ were found to show minima, and then
to give rise to stable molecular configurations.

An error in such calculations, related to the transformation from parabolic coordinates (useful in treating hydrogen atoms far apart) to elliptic ones (used for nearly united atoms), was corrected some months later by Lennard-Jones \cite{LJ}, while other calculations of excited states for H$_{2}^{+}$, performed by following Wilson's approach \cite{wilson1}, were carried out by Teller \cite{teller1}, who considered the molecular ion H$_{2}^{+}$ as composed of H and H$^{+}$ with H in its ground or first excited state. Teller found that only one possible state -- the $3d\sigma $ state -- corresponded to a stable configuration, and provided numerical values for the equilibrium distance and the heat of dissociation, along with a study of the band spectrum.

\subsection{H$_2$ excited states and ionic structures}

Quantitative studies on the ground state of the H$_2$ molecule were performed, as we have seen above, by Condon \cite{condon2}, while Pauling \cite{pauling5} considered also its excited states, assumed to be obtained by means of a normal hydrogen atom and a hydrogen atom in various excited states. Because of a very small inter-electronic interaction, the properties of the molecule turned out to be, at a good approximation, those of H$_{2}^{+}$, and Pauling indeed realized that the system effectively reduced to that of the molecular ion with an electron in an outer orbit, so that it could be considered a polar compound of H$^{+}$ and H$^{-}$. The corresponding zeroth order wavefunction,
\begin{equation}
\frac{1}{\sqrt{2+2S^{2}}}\left( \varphi _{1}\varphi _{2}+\psi _{1}\psi
_{2}\right) ,  \label{d11}
\end{equation}
clearly represents a polar state,\footnote{Contrary to what introduced by Heitler and London (see Eq. (\ref{p9}), here {\it only} $\psi_i \psi_j$ or $\varphi_i \varphi_j$ terms are present, while mixed terms of the Heitler-London form, $\psi_i \varphi_j$, are absent.} but, unfortunately, Pauling did not obtain reliable results with first order perturbation theory.

Just the contrary, instead, happened with the calculations performed by Hylleraas \cite{hylleraas1}, who followed the same reasoning as Condon \cite{condon2} in order to get eigenvalues and eigenfunctions of the H$_2$ molecule starting from solutions of the two-center problem of H$_2^+$. His novel idea to introduce ``fractional nuclear charges for the outer electron'' led to a very small perturbation for excited states, so that just a first order perturbative calculation of the eigenfunctions was enough to obtain very accurate results for both small and intermediate internuclear separations in H$_2$.

Independently of Pauling, ionic structures were successfully introduced into homopolar molecules by E. Majorana \cite{EM2} in 1931, inspired by some apparently conflicting spectroscopic evidences about excited states of the hydrogen molecule. The puzzling phenomenon was indeed observed, concerning the decay of the excited $\left( 2p\sigma \right) ^{2}$$\, ^{1}\Sigma _{g}$ (gerade) state into the (ungerade) $1s\sigma$$\, 2p\sigma $$\, ^{1}\Sigma _{u}$ state in the infrared spectral region \cite{weizel2}. The theoretical justification of even the existence of the $\left( 2p\sigma \right) ^{2}$$\, ^{1}\Sigma _{g}$ term, along with an explanation of its abnormal energy level, when compared to similar atomic systems, urged a reconsideration of the Heitler-London paradigm. Here, only configurations corresponding to one electron in each atom of the molecule were considered, but, in the theoretical description of the H$_2$ molecule, Majorana included different configurations where both electrons or no electron belong to a given atom. Thus, in addition to the Heitler-London chemical reaction H$+$H$\, \leftrightarrow \, $H$_{2}$, Majorana considered also the reaction H$^{+}+$H$^{-} \leftrightarrow \, $H$_{2}$, which contains ionic structures, and was able to take into account the formation of the $\left( 2p\sigma \right) ^{2}$$\, ^{1}\Sigma _{g}$ state in the H$_2$ molecule.
\begin{quote}
This does not mean, however, that it is a polar compound since, because of the equality of the constituents,
the electric moment changes sign with a high frequency (exchange frequency) and therefore cannot be observed. It is in this sense that we speak of a pseudopolar compound \cite{EM2}.
\end{quote}
Majorana then succeeded to prove the existence of a stable molecular state with both electrons in excited $2p$ orbitals, and the numerical results for the equilibrium internuclear distance of the molecule in the relevant state and for the corresponding vibrational frequency resulted to be amazingly close to the experimental
observations \cite{nostro2}.

\subsection{He compounds and three-electron bond}

The formation of the He$_{2}$ molecule, as we have seen above, was successfully considered for the first time by Heitler and London \cite{heitler1}, who concluded that no stable molecule of such a kind could be formed (in the normal state), due to the Pauli principle. The problem then arose to study the possible formation of other compounds involving helium, this really requiring the introduction of new concepts in order to fully understand the nature of the chemical bond involved and its formation.

In 1930, by following the Heitler-London method, G. Gentile \cite{gentile1} computed the interaction energy between H and He as well as between two helium atoms in their ground state; his perturbative calculations (which included, at second order, polarization forces) showed that normal He and H do not form a stable molecule.

One year later, Pauling reconsidered the general problem \cite{pauling2} (see also \cite{pauling4}), and  looked for approximate solutions of the Schr\"odinger equation for a diatomic system built of a pair of electrons attached to one nucleus and a single electron attached to the second nucleus, recognizing that resonance forces corresponding to the exchange of three electrons were mainly repulsive. This allowed him to propose an exchange mechanism for the occurrence of a {\it three-electron bond} that was a direct generalization of what originally introduced by Heitler and London, where the resonance phenomenon involved degenerate (or nearly degenerate) electronic states, and a rule was formulated:
\begin{quote}
A three electron bond, involving one eigenfunction for each of two atoms and three electrons, can be formed in case the two con\-fi\-gu\-ra\-tions A$: \! \cdot$B and A$\cdot \! :$B correspond to essentially the same energy \cite{pauling2}.
\end{quote}
This applied -- Pauling suggested -- to the cases of He$_2^+$, NO, NO$_2$, O$_2$, etc. Instead, in the presence of four electrons, he deduced that there is no tendency to form a strong molecular bond, because two of them are necessarily nuclear symmetric while the remaining two are nuclear antisymmetric.

\subsection{The formation of the helium molecular ion}

The problem of the formation of He$_2^+$ was considered only qualitatively by Pauling, by assuming  -- following Weizel \cite{weizel3} -- that its formation was due to the bonding of a neutral helium atom with a ionized one, He $+$ He$^+ \rightarrow$ He$^+_2$, in full analogy with the case of the hydrogen molecular
ion H$^+_2$ (but with three electrons instead of only one), rather than with that of the compound HeH, which exhibits a three-electron bond. A similar process was envisaged also for the (excited) neutral helium molecule: there, an unpaired $1s$ electron comes into play as a result of the excitation of one atom, whose interaction with the pair of $1s$ electrons of the other atom eventually leading to a three-electron bond. The contribution to the chemical bond from the remaining outer electron could be neglected, so giving rise to a
hydrogen-like spectrum.

Later in 1933, Pauling did provide a quantitative treatment of the normal state of He$_{2}^{+}$ by means of a variational method \cite{pauling3} with a wavefunction for the system of two helium nuclei and three electrons in terms of hydrogen-like $1s$ functions, by assuming each electron to interact only with one nucleus. Minimization with respect to the effective nuclear charge $Z$ ($Z_{\rm eff}=1.833$) resulted into fairly accurate numerical estimates for the equilibrium internuclear distance ($R_{\rm eq}=1.085 \, {\rm A}$), the dissociation energy ($D_{{\rm He}_2^+}=2.47 \, {\rm eV}$) and, to a lesser extent, the vibrational frequency ($\omega \sim 1950 \, {\rm cm}^{-1}$).

By adopting the same variational procedure, and with similar component wavefunctions, approximately the same numerical results were obtained much earlier by Majorana \cite{EM1}, who (differently from Pauling) was able to construct the appropriate eigenfunctions of the system in accordance with its symmetry properties, by using a typical group theory reasoning. The relevance of {\it inversion} symmetry -- the total electronic wavefunction must show a definite symmetry with respect to the midpoint of the internuclear line -- was correctly emphasized, thus allowing him to show that two molecular states are possible for the He$_2^+$ molecular ion, only one of which corresponding to the bonding molecular orbital of the ion, such a configuration just reflecting the fact that the ground state of the system exhibits resonance between the He$: \! \cdot$He and He$\cdot \! :$He configurations.

\section{Intermolecular interactions}
\label{intermol}

\noindent Simultaneously to the development of appropriate methods able to explain satisfactorily the homopolar valence in molecules, the building up of an accurate quantum theory of intermolecular forces was pursued starting from 1930, which further deepened the understanding of the nature of the chemical bond.

A first step towards this direction was carried out by R. Eisenschitz and London \cite{london3}, who devised a general perturbative approach that included, in a unified scheme, both homopolar and van der Waals forces. The underlying idea was to recognize the homopolar valence according to quantum mechanics as the result of a saturation mechanism:
\begin{quote}
This saturation mechanism is of the type where, for instance, a single-valued atom cannot bind to a third atom with the same force with which it binds to a second atom \cite{london3}.
\end{quote}
Since a bound valence can be broken only by supplying to the molecule an amount of energy that coincides with the chemical activation energy, the valence forces decrease exponentially with the distance between the constituent atoms, $\sim \erm^{-\alpha R}$, and were thus identified as relatively short ranged.
Instead, since for the polarization forces ``no saturation is present, they are not related to the chemical interactions of neutral atoms, but to the van der Waals attraction forces'' \cite{london3}, and are thus based  on the interaction between induced dipoles. They were found to decrease with the interatomic distance as $\sim {1}/{R^{7}}$, so that they vanish more slowly at infinity than the homopolar binding forces, and then  dominate at large distances. Just on the basis of such considerations, Eisenschitz and London \cite{london3} developed a general procedure for successive approximations of the atomic interactions, which included valence as a first order effect, and then van der Waals forces.

Few months later, London also considered medium and long range forces \cite{london4} \cite{london5} in order to characterize the intermolecular attraction playing a role in the phase transition from liquid to gas. Within the old quantum theory, van der Waals attraction was found to be proportional to the square of the quadrupole moment of the system, so that a huge moment had to be assumed for noble gases to fit with observations. However, with the advent of the new quantum mechanics, it became clear that such quadrupole moments are extremely small (or vanishing at all) due to the spherical symmetry of the system, and thus ``there must be other, and essentially stronger, forces which are not related exclusively to the existence of a molecular quadrupole" \cite{london4}. From Eisenschitz and London, it was in fact known that the attractive potential between spherical symmetric H atoms at large distances goes as $\sim {1}/{R^{6}}$, so that it is certainly predominant over that derived for polarizable quadrupole molecules, scaling as $\sim {1}/{R^{8}}$.

Indeed, molecular attraction was determined as a second order perturbative effect, both in the case of atomic noble gases and for polyatomic molecules, even if such molecules have more complex symmetry properties (spherical symmetry holds only in the lowest rotational state, with $J=0$). The second order perturbation energy, describing the interaction between a molecule in the state $k$ and another molecule in the state $\ell$ is given by
\begin{equation}
\mathcal{E}_{k\ell}^{( 2) }=\sum_{k^{\prime },\ell^{\prime}} \frac{\left| V_{k \ell,k^\prime \ell^\prime}\right| ^{2}}{ E_{k}+E_{\ell}-E_{k^\prime}-E_{\ell^\prime}},  \label{h1}
\end{equation}
where the terms in the summation with vanishing denominator had to be left out. The interaction energy matrix $V_{k \ell,k^\prime \ell^\prime}$, depending explicitly on the interatomic distance $R$, was
expanded in terms of ${1}/{R}$ as:
\begin{equation}
V_{k \ell,k^\prime \ell^\prime} =\frac{\mu _{k,k^\prime}\mu_{l,\ell^\prime}}{R^{3}}P_{3}+\frac{\tau _{k,k^\prime}\mu_{l,\ell^\prime}}{R^{4}}P_{4}+\frac{\mu _{k,k^\prime}\tau
_{l,\ell^\prime}}{R^{4}}P_{4}^{^{\prime }}+\frac{\tau _{k,k^{^{\prime
}}}\tau _{l,\ell^\prime}}{R^{5}}P_{5},  \label{h2}
\end{equation}
where $\mu _{k,k^\prime}$, $\tau _{k,k^\prime}$ are dipole and quadrupole moments, respectively, and  $P_{3}$, $P_{4}$, \dots  \ are functions (of order $1$) of the spatial orientation of the multipole. The main contribution to the interaction energy in (\ref{h1}) comes just from the first term in Eq. (\ref{h2}), which is then proportional to ${1}/{R^{6}}$, as already found by Eisenschitz and London, while the others give subdominant contributions with ${1}/{R^{8}}$, \dots \ scaling.

The intermolecular interaction (\ref{h1}) (whose range of validity is restricted to $\left| V_{k\ell,k^\prime \ell^\prime}\right| \ll | E_{k}+E_{l}-E_{k^\prime}-E_{\ell^\prime}| $) was decomposed into four terms as follows \cite{london4}:
\begin{equation}
\mathcal{E}_{k\ell}^{( 2) }=\mathcal{E}_{k\ell}^{( rr) }+ \mathcal{E}_{k\ell}^{( rg) }+\mathcal{E}_{k\ell}^{( gr) }+ \mathcal{E}_{k\ell}^{( gg) }  . \label{h3}
\end{equation}
The contribution $\mathcal{E}_{kl}^{( rr) }$, corresponding to the interaction between rigid and rotating multipoles, was envisaged as a directional effect, while the term $\mathcal{E}_{kl}^{( rg) }+\mathcal{E}_{kl}^{(gr) }$ was identified with the Debye induction effect, due to the polarizability of the slowly rotating molecule in a quasi-static field, and, finally, the contribution $\mathcal{E}_{kl}^{( gg) }$ (driving a dispersion effect) was recognized to be due to the fast periodic influence of the motion of inner electrons on each other, giving rise to the temperature independent part of the van der Waals attraction.

The origin of intermolecular forces was further discussed by London \cite{london5}, later in the same year 1930, also by means of a simplified but exactly soluble model, in order to evaluate sublimation and adsorption heats, dissociation energies of molecules interacting via van der Waals forces, and so on. This was achieved by considering two quasi-elastic dipoles with variable electric moment $e\mathbf{r}$, mass $m$, momentum $\mathbf{p}$ and polarizability $\alpha$, put at a fixed distance $R$, whose Hamiltonian is
\begin{equation}
H=\frac{1}{2m}\left( \mathbf{p}_{1}^{2}+\mathbf{p}_{2}^{2}\right) +\frac{e^{2}}{2\alpha }\left( \mathbf{r}_{1}^{2}+\mathbf{r}_{2}^{2}\right) +\frac{e^{2}}{R^{3}}\left( x_{1}x_{2}+y_{1}y_{2}-2z_{1}z_{2}\right) .  \label{h4}
\end{equation}
The second term modeled the quasi-elastic potential energy of the individual mo\-le\-cu\-les, while the third one is the interaction potential of the molecules with each other. By introducing the normal coordinates of the principal vibrations it was possible to cast it into an Hamiltonian of six decoupled oscillators, whose
total energy was easily found to be:
\begin{eqnarray}
E &=&h\nu _{0}\left(
n_{X}^{+}+n_{Y}^{+}+n_{Z}^{+}+n_{X}^{-}+n_{Y}^{-}+n_{Z}^{-}+3\right)  \notag
\\
&&+\frac{h\nu _{0}}{2}\frac{\alpha }{R^{3}}\left(
n_{X}^{+}+n_{Y}^{+}-2n_{Z}^{+}-n_{X}^{-}-n_{Y}^{-}+2n_{Z}^{-}\right)
\label{h5} \\
&&+\frac{h\nu _{0}}{2}\frac{\alpha ^{2}}{R^{6}}\left(
n_{X}^{+}+n_{Y}^{+}+4n_{Z}^{+}+n_{X}^{-}+n_{Y}^{-}+4n_{Z}^{-}+6\right) + \dots ,
\notag
\end{eqnarray}
$\nu _{0}={e}/{\sqrt{m\alpha }}$ being the eigenfrequency of the isolated dipole. Here the first term, independent of $R$, gives the eigenenergy of the two molecules taken together, while the second one, depending on the resonance degeneracy, leads to dipole type forces. The third term describes the interaction between non-excited molecules, which is always attractive; for the lowest state with $n_{X}^{+}=n_{Y}^{+}=...=n_{Z}^{-}=0$, it reduces to the zero point contribution
\begin{equation}
\varepsilon =-\frac{3}{4}\frac{h\nu _{0}\alpha ^{2}}{R^{6}} \, .  \label{h6}
\end{equation}
Interestingly, such a result led London to link the appearance of forces between molecules in their normal state to the presence of a zero-point motion, subject to the restriction $R^{3} \gg \alpha $.

\section{Concluding remarks}

\noindent At the end of his review paper, in 1929 van Vleck hopefully asked:
\begin{quote}
Is it too optimistic to hazard the opinion that this is perhaps the beginnings of a science of ``mathematical chemistry" in which chemical heats of reaction are calculated by quantum mechanics just as are the spectroscopic frequencies of the physicist? Of course the mathematics will be laborious and involved, and the results always successive approximations. The theoretical computer of molecular energy levels must have a technique comparable with that of a mathematical astronomer \cite{vanvleck}.
\end{quote}
What discussed above has shown that the quantum explanation of the nature of the chemical bond in  molecules was (unfortunately) not at all only a matter of laborious and involved mathematical calculations, but its story has been rather intricate, when compared to that of the analogous description of atoms. A number of authors contributed with more or less significant results, by starting (as for atomic physics) with the classification and ordering of spectroscopic data.

The distinguishing work performed by Hund (then followed by many others) served to establish molecular spectroscopy as an essential tool for developing a quantum theory of the chemical bond, but, even with the aid of group theory, the first steps along this line were still oriented by the framework of the old quantum theory, or -- in other terms -- they focused too much on the methods typical of atomic physics. However, it was soon recognized that such methods were at variance with those required for a successful description of molecules, and  approximation schemes were developed in order to find reliable solutions to the Schr\"odinger equation of the given non-central molecular problem (diatomic molecules). A quantitative theory accounting for rotational, vibrational and electronic terms of a molecule was developed by Born and Oppenheimer, showing that the corresponding energy contributions for those states could be obtained systematically as successive terms in a suitable series expansion involving the electron-to-nuclear mass ratio, thus becoming possible to neglect the nuclear motion at a first approximation.

Novel ideas, however, had to be introduced in order to gain more physical insight into the molecular problem, and Heitler and London succeeded to translate Heisenberg's resonance mechanism (introduced in atomic physics) into a suitable form able to account for the stability of the hydrogen molecule in terms of genuinely quantum exchange forces. They also predicted the dependence of the total electronic energy on the internuclear distance, and showed, remarkably, that no stable He$_2$ may be formed, thus opening a viable route towards predictive quantum computations in molecular physics.

Heitler and London's approach originated from the concept of chemical valence, and, in a sense, it was aimed to explain the chemical bond by using properties of the constituent atoms, such atoms being assumed to retain their properties in the molecule to a large extent. People frequently accustomed to work with molecular spectroscopy, however, were used to observe similar properties in many different molecules (formed with different atoms) with the same number of electrons, and it was then quite natural to assume a molecule to be a collection of nuclei fixed in given spatial positions with an additional electron cloud surrounding them. This was at the roots of the idea of  ``promoted electrons'' introduced by Mulliken, and just within such a framework (followed by several other spectroscopy-based scientists) the concept of molecular orbitals occupied by each electron appeared by means of Lennard-Jones. A molecular orbital spreads over the whole of the molecule, so that the chemical bond resulted -- in this view -- from ``sharing'' electrons among the constituent atoms, these then loosing their identity to a large extent, as opposed to the Heitler-London method.

These first steps allowed the thorough (quantitative) description of the most simple molecules, H$_2^+$ and H$_2$, as well as the prediction of the non existence of a stable helium molecule, but the (qualitative) description of more complex molecules required the formulations of additional key concepts. One-electron and three-electron bonds were introduced by Pauling, in order to generalize the Heitler-London resonance phenomenon to molecules with a number of valence electrons different from two, and the notion of {\it directed valence} later elaborated by J.C. Slater \cite{slater1} and Pauling \cite{pauling1} was just an example (among many others which appeared subsequently) of how it is possible to manage complex molecular problems.

Two further important results were, instead, achieved by Majorana (and, independently, by Pauling). He generalized the Heitler-London theory to include ionic structures in homopolar compounds: exchange and polar interaction terms were combined in a single wavefunction with different weights, in order to properly describe the different contributions leading to the formation of a stable chemical bond. In a sense, this realized a bridge between Mulliken-inspired molecular orbitals  (with their characteristic polar terms) and Heitler-London's valence theory. Majorana also succeeded to prove theoretically the stability of the helium molecular ion, by extracting the appropriate symmetry properties of the molecular system and then building a suitable quantum theory within the conceptual framework of Heitler-London's.

Finally, the study of intermolecular interactions, including van der Waals forces, further enlightened the subject, by allowing London to elucidate the role of zero-point motions.

Although a number of other authors have contributed significantly to the development of quantum molecular physics, with results that have delineated and sharpened its subsequent shape, what discussed here remains the basic pillars of the theory, and we hope that the present paper has concurred to disentangle the intricate genesis of the quantum theory of the chemical bond.

\end{document}